\documentclass[10pt, draftcls, onecolumn, journal]{IEEEtran}
\usepackage{balance}
\usepackage{cite, graphicx,amssymb,color,  pict2e, multirow, rotating} 
\usepackage{graphicx,amssymb}
\usepackage[tbtags]{amsmath}
\usepackage{times,amsmath}
\usepackage{subfig}
\usepackage{flushend}
\usepackage{amsmath}
\usepackage{epsfig}
\usepackage{amssymb}
\usepackage{hyperref}
\usepackage{color}
\usepackage{psfrag}
\usepackage{subfig}
\usepackage[usenames,dvipsnames]{xcolor}
\usepackage{textcomp}
\usepackage{array}
\usepackage{tikz}
\usetikzlibrary{matrix,decorations.pathreplacing}

\usepackage{tabularx,ragged2e}
\newcolumntype{C}{>{\Centering\arraybackslash}X} 

\usepackage{algorithmic}
\usepackage{algorithm}

\begin{document}
\vspace{-12mm}
\title{Contention Intensity based Distributed Coordination for V2V Safety Message Broadcast}
\vspace{-3mm}

\author{\IEEEauthorblockN{Jie~Gao, \IEEEmembership{Member, IEEE},   
		Mushu~Li, \IEEEmembership{Student Member, IEEE},
		Lian~Zhao, \IEEEmembership{Senior Member, IEEE}, 
		and Xuemin~(Sherman) Shen, \IEEEmembership{Fellow, IEEE} 
	}
	\thanks{
		
		J. Gao and L. Zhao are with the Department of Electrical and Computer Engineering, Ryerson University, Toronto, ON, M5B 2K3, Canada (e-mail: \{j.gao, l5zhao\}@ryerson.ca).
		
		M. Li and X. Shen are with the Department of Electrical and Computer Engineering, University of Waterloo, Waterloo, ON, N2L 3G1, Canada (e-mail: \{m475li, sshen\}@uwaterloo.ca). 
	}
}

\maketitle

\begin{abstract}

In this paper, we propose a contention intensity based distributed coordination (CIDC) scheme for safety message broadcast. By exploiting the high-frequency and periodical features of the safety message broadcast, the application-layer design of the CIDC enables each vehicle to estimate the instantaneous channel contention intensity in a fully distributed manner. With the contention intensity information, the MAC layer design of CIDC allows vehicles to adopt a better channel access strategy compared to the 802.11p. This is because CIDC selects the initial back-off counter for each new packet deterministically, i.e., based on the contention intensity, instead of randomly. The proposed CIDC is modeled, and key performance indicators in terms of the packet collision probability and average contention delay, are derived. It is shown that the proposed change in the initial counter selection leads to a system model completely different from the classic Markov chain based model. Moreover, the proposed CIDC, fully distributed and compatible with the 802.11p, can achieve both a much lower collision probability and a smaller contention delay compared with 802.11p at the cost of a small communication and computation overhead. Extensive simulation results demonstrate the effectiveness of the CIDC in both of the accurate and the erroneous contention intensity estimation scenarios.

\end{abstract}

\vspace{-0.2cm}

\IEEEpeerreviewmaketitle

\vspace{-3mm}
\section{Introduction}\label{s:intro}

Vehicle-to-vehicle (V2V) communications are a cornerstone of connected vehicles (CV), which are emerging as an important component of the next generation intelligent transportation systems (ITS) \cite{S_NLU2014}. The deployment of connected vehicles, combined with automated driving, is expected to significantly reduce traffic accidents and the resulting economic loss through integrating communications including V2V, vehicle-to-infrastructure (V2I), etc. and enabling an awareness of the surrounding traffic environment and events at all vehicles \cite{R_KKockelman2016}. As an effort of deploying CV, technologies and standards have been actively developed. Dedicated short-range communications (DSRC) have been tested as an enabling technology for V2V and V2I communications \cite{S_JKenney2011}. \textcolor{Black}{The DSRC standard adopts seven channels, i.e., one dedicated control channel for safety applications and six service channels for non-safety applications.} The IEEE 802.11p standard covers the physical layer and media access control (MAC) layer designs of DSRC. The IEEE 1609 standard stack addresses the MAC, transport, and network layers. The IEEE 802.11p and IEEE 1609 collectively form the wireless access in vehicular environments (WAVE) standards while the SAE J2735 standard defines the message set to be utilized in the application layer.

In the V2V communications for CV applications, a particularly, if not the most, important component is the broadcast of safety messages. Such broadcast corresponds to the Basic Safety Message (BSM) in the SAE J2735 standard in the US  or the Cooperative Awareness Message (CAM) in the ITS standard of European Telecommunications Standards Institute~\cite{J_BKloiber2016}. The safety messages are single-hop, periodical (i.e., time-driven as opposed to event-driven), and carry safety-related status information of vehicles such as their speed, acceleration, position, and direction. Through the broadcast of the safety messages, vehicles can be aware of each other's status, and traffic accidents can be reduced. As a result, the information conveyed in the safety message broadcast is the foundation to support all V2V safety applications \cite{S_JKenney2011}. Meanwhile, safety messages need to be exchanged at a high frequency, e.g., 10 messages per second, to be able to support safety applications in CV. Such a high frequency renders the safety message broadcast the major data traffic load on the DSRC control channel. Therefore, the safety message broadcast is significant in V2V communications in terms of both its importance and its data traffic volume. Accordingly, a protocol for V2V communications should incorporate elaborate designs to support reliable safety message broadcast.

In general, a protocol design for V2V communications faces a fundamental trade-off: a better performance is usually achieved at the cost of a larger overhead \cite{J_MHadded2015}. For the safety message broadcast, the 802.11p has the advantage that it is fully distributed, requires no coordination, and yields no overhead. However, the channel contention design of 802.11p suffers from a high collision probability under a high network load \cite{Y_Bi2016}. Thus, the 802.11p and most works built on it can be considered as designs that trade off performance improvement potentials for a small overhead. Other works in the literature, as will be discussed in detail in Section~\ref{s:Related}, aim to achieve a substantially improved performance at the cost of more coordination and a larger overhead. 

The objective of this work is to develop a distributed scheme for safety message broadcast that improves the performance substantially compared to the 802.11p at the cost of a small overhead. In order to achieve this target, we exploit the unique features of the safety message broadcast, i.e., high message frequency, periodicity, and uniform format, and develop our designs based on these features. The contributions of this work are as follows. 


First, we propose a novel contention intensity based distributed coordination (CIDC) scheme to improve the performance of safety message broadcast. 
The proposed design is fully distributed and compatible with 802.11p, features a small communication and computation overhead, and achieves a substantial performance improvement as compared to 802.11p.




Second, the proposed CIDC is modeled and characterized, and insights regarding the contention delay and the packet collision are found. 
Analytical results on the contention delay and collision probability of the CIDC are derived. It is demonstrated that a seemingly simple modification of the 802.11p MAC by the proposed CIDC leads to a very different system model and a different underlying cause of packet collision compared to the 802.11p. 

Third, the performance of the proposed CIDC is demonstrated using extensive simulations. The collision probability and average delay of the CIDC are compared to both the analytic results and those of the 802.11p MAC. It is shown that the CIDC can substantially reduce both the collision probability and the average contention delay in a wide range of vehicle density compared to the 802.11p MAC even when errors are introduced to account for vehicle mobility and other potential practical concerns. 

\section{Related Works}\label{s:Related}

This section reviews the existing works related to safety message broadcast in V2V communications in the literature. We categorize the related works into two classes based on their relation with the 802.11p.

The first class of works focuses on the performance analysis of the 802.11p protocol, among which some studies propose various parameter adaption schemes for the performance optimization of the 802.11p. The analysis on the performance of 802.11p can be found in \cite{L_CCampolo2011} - \cite{J_KHafeez2016}. Han \textit{et al.} proposed an analytical model for the performance of 802.11p MAC in \cite{J_CHan2012} with a focus on different access categories. Ma \textit{et al.} developed a characterization of the reliability of safety message broadcast using different transmitter-centric and receiver-centric metrics \cite{J_XMa2011}. Ye \textit{et al.} analyzed the broadcast efficiency and reliability trade-off in 802.11p in \cite{J_FYe2011} and proposed performance optimization based on adapting the contention window size or storing the message with a certain probability. Fallah \textit{et al.} studied the performance of broadcast in 802.11p in a highway scenario and proposed congestion control based on adjusting the transmission power and message frequency \cite{J_YFallah2011}. Hafeez \textit{et al.} derived a performance analysis considering the mobility of vehicles and proposed performance enhancement based on adapting the message frequency, transmission power, and contention window size \cite{J_KHafeez2013}. The above works based on adapting the parameters of the 802.11p aim to achieve a performance improvement while sustaining a distributed structure and small overhead. However, the approach based on parameter adaption may encounter limitations in practice. For instance, adapting the message frequency is not always a feasible option as the requirement on the message frequency for supporting safety applications can be too stringent to be practically adjustable. As for adapting the contention window size, it essentially introduces a trade-off between the delay performance and the collision performance instead of improving both. There are also research works built on the 802.11p that are not based on parameter adaption~\cite{J_XMa2012} - \cite{J_YPark2013}. Wu \textit{et al.} proposed a deterministic channel access for safety message broadcast based on protocol sequences \cite{J_YWu2014}. Park and Kim developed an application layer design based on a random offset in BSM epoch selection to improve the MAC layer performance of 802.11p when the message frequency is not adaptable~\cite{J_YPark2013}. 

Recognizing the performance limitation of the 802.11p under a high network load, the second class of works aims to develop new protocols instead of improving the 802.11p. Various MAC protocols for vehicular communications have been proposed, including both distributed and centralized and both contention-based and contention-free designs \cite{J_MHadded2015}. 
Omar \textit{et al.} proposed a distributed time division multiple access (TDMA)-based MAC Protocol in the scenario of bi-directional traffic flow with vehicles equipped with two transceivers \cite{J_HOmar2013}. The protocol can significantly decrease the collision probability and improve throughput as compared to 802.11p. As TDMA based MAC lacks scalability with respect to the number of vehicles, Lyu \textit{et al.} designed a slot sharing TDMA based MAC for safety message broadcasting \cite{JF_Lyu2018}. Bharati \textit{et al.} proposed a distributed cooperation based MAC in \cite{J_SBharati2013}, in which vehicles form clusters and a cooperation header is inserted into a packet when a vehicle decides to help others. Zhang \textit{et al.} proposed a centralized TDMA based MAC, in which a roadside unit (RSU) is needed to coordinate the communications of the vehicles \cite{J_RZhang2015}. A TDMA based MAC is also proposed in \cite{J_SBharati2016} for relaying broadcast messages. For a scenario in which multiple channels are available and each vehicle has two radio heads, Almotairi developed a frequency hopping based MAC \cite{J_KAlmotairi2015}. Ye and Zhuang proposed an adaptive control MAC solution that switches between 802.11p and TDMA-based MAC depending on the network load \cite{J_QYe2016}. A study of the periodical broadcast with  geolocation-based access in LTE V2X can be found in \cite{J_FMV2018}. It can be seen that these works usually aim to achieve a significant performance improvement at the cost of a larger overhead necessary for implementing an effective coordination in the network.

This work can be categorized into the first class. However, instead of adapting the parameters in the 802.11p, we aim to achieve a significant performance improvement by exploiting the unique features of the safety message broadcast and developing the designs accordingly.

\section{The Proposed Design}\label{s:mac}

The proposed CIDC exploits three unique features of safety message broadcast, as summarized in Table~\ref{T:comp}. 

First, while the high message frequency can lead to congestion, it enables vehicles to update their information on surrounding vehicles timely. Moreover, a high message frequency also implies that the impact of vehicle mobility in the duration of a message cycle is very limited.~\footnote{With a 10Hz message frequency, the distance between two vehicles driving toward each other at 60km/h reduces by only 3 meters in a message cycle.} Consequently, the impact of the mobility on the topology of the vehicular network within several message cycles can be neglected.

Second, as the safety messages are periodical, there is a regular pattern regarding the message arrival instants of each vehicle, which can be observed by other vehicles through message exchanges. Such observations can be exploited to improve the channel access strategy of the vehicles. 

Third, as the safety message has a uniform format and size, there is a regular pattern regarding the packet transmission duration, which can be exploited in the protocol design and modeling.   

It should be noted that the CIDC is dedicated to the time-driven safety message broadcast. While other types of messages are not covered, the proposed design can be used as a building block in a comprehensive protocol that covers all types of V2V communications. For example, a protocol can adopt the CIDC for the broadcast on the DSRC control channel and the 802.11p for the unicast on the service channels. 

\begin{table}
	\begin{center}
		\caption{Features of Safety Message}\label{T:comp}
		{\setlength{\extrarowheight}{1.5pt}
			\begin{tabularx}{0.76\textwidth}{|C|C|}\hline
				\textbf{Feature} &  \textbf{Implication}  \\ \hline
				High Message Frequency & Information updated timely; Mobility not a major concern.  \\ \hline
				Time-Driven (Periodical) & Regularity in the message arrival instants \\ \hline
				Uniform in Format and Size & Regularity in the transmission duration
				\\ \hline
			\end{tabularx}
		}
	\end{center}
	\vspace{-0.3cm}
\end{table}

\subsection{Assumptions}

The following assumptions are made based on the features of safety messages:
\begin{itemize}	
	\item Given the high frequency of safety message broadcast (e.g., $\lambda = 10$ messages per second), the variation in the set of neighbor vehicles due to vehicle mobility is negligible during several message periods.
	\item Vehicle $i$ has a unique random offset $\sigma_i$ independently and uniformly drawn from $[0, 1/\lambda)$. This can prevent messages of different vehicles from arriving and contending for channel access at the same instant. The offset $\sigma_i$ can be fixed for each vehicle or refreshing at a frequency $\mu$ such that $\mu \ll \lambda$. In either case, the random offset for each vehicle is constant for a significant number of message cycles. 
	\item A new safety message expires and replaces an existing safety message that has not been sent.
	\item The vehicles have the capability of processing real-time information, which is necessary to support any safety applications.
\end{itemize}
To facilitate the modeling and analysis of the CIDC, the following assumption is also used:
\begin{itemize}	
	\item Safety messages have a uniform size. Specifically, the duration of a message transmission plus a DCF Interframe Space (DIFS) is assumed to be a multiple of the duration of a time slot, i.e., $T_\mathrm{Tx}+ T_\mathrm{DIFS} = K T_\mathrm{s}$, where $T_\mathrm{Tx}$, $T_\mathrm{DIFS}$, $T_\mathrm{s}$ are the length of a message transmission, the length of a DIFS, and the length of a time slot in the DCF, respectively.
\end{itemize}

\subsection{CIDC: the Model}

The term \textit{contention intensity} refers to the number of safety messages that are either waiting for channel access or currently transmitting at a given time instant. The CIDC consists of two parts. The first part is \textit{distributed contention intensity estimation}, an application layer function that facilitates information exchange for vehicles to estimate the channel contention intensity. The second part is the \textit{access strategy}, a MAC layer design built on the DCF of the 802.11p MAC but modified to exploit the contention intensity information. 

\begin{figure}[!t]
	\begin{center}
		\includegraphics[angle=0,width=0.65\textwidth]{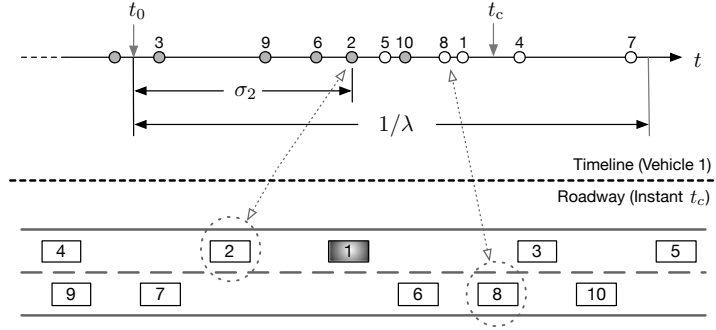}
		\caption{Illustration of contention intensity estimation.}
		\label{f:MeasureCK}
	\end{center}
	\vspace{-4mm}
\end{figure}

The application-layer design for the distributed contention intensity estimation is as follows.
\begin{itemize}
\item Vehicle $i$ includes its offset $\sigma_i$ in its safety messages.
\item Vehicle $i$ extracts the offsets $\sigma_j$ of vehicles $\{j| j\neq i\}$ within its range from received safety messages. 
\item Vehicle $i$ estimates the contention intensity at instants when a new safety packet is generated at its MAC layer.~\footnote{A MAC layer packet contains an application layer message in its payload. In the rest of the paper, `packet' and `message' are used interchangeably when there is no confusion.} 
\end{itemize}
The estimation of contention intensity is illustrated in Fig.~\ref{f:MeasureCK}. In this example, there are 10 vehicles within each other's communication range, shown in the bottom part of Fig.~\ref{f:MeasureCK}. The perspective of vehicle 1 is used as an example. As shown in the top part of Fig.~\ref{f:MeasureCK},  vehicle 1 maintains a timeline and marks the instants at which each vehicle in its communication range generates a safety message (shown as the hollow circles on the timeline in Fig.~\ref{f:MeasureCK}) based on the safety messages received in previous cycles. When a new safety message is received from a neighbor vehicle, the corresponding mark is changed (shown as solid circles on the timeline) to indicate that the message is no longer contending for channel access. Then, the messages that have been generated by neighbors and not yet received in the interval from the beginning of this message cycle, marked as $t_\mathrm{0}$, till the current time instant, marked as $t_\mathrm{c}$, are contending for channel access. Counting the number of such messages gives the instantaneous contention intensity, which is 3 in the example given by Fig.~\ref{f:MeasureCK}. \textcolor{Black}{Note that the beginning of message cycles should be aligned at all vehicles. For example, the beginning can be determined based on the GPS time.}
 
\begin{figure}[!t]
	\subfloat[Illustration of initial back-off counter determination -  entering from the end.]
	{\includegraphics[angle=0,width=0.5\textwidth]{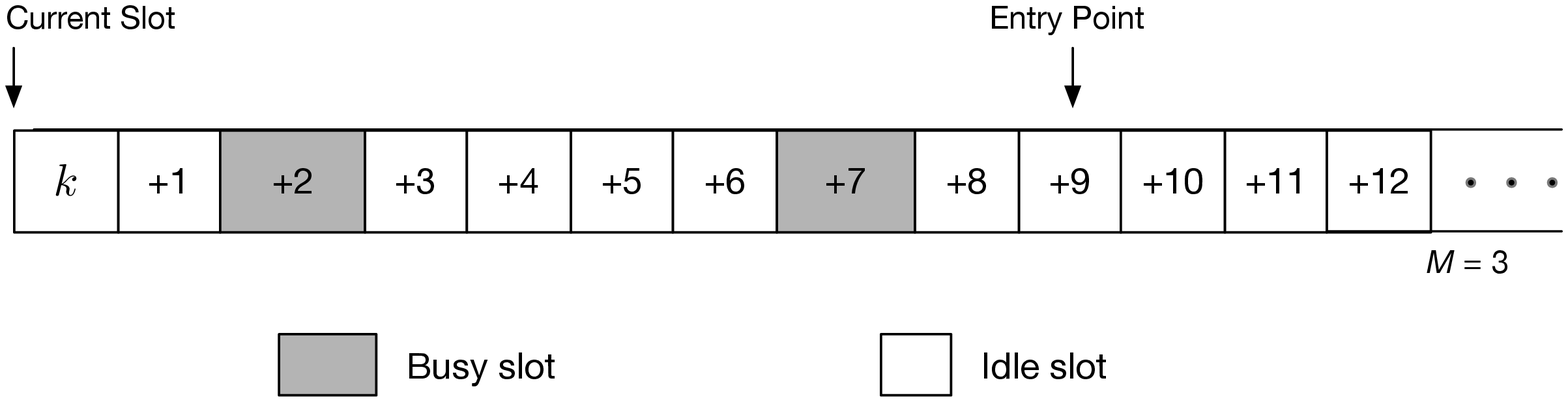}
		\label{f:EntranceP1}}
	\vspace{1mm}
	\subfloat[Illustration of initial back-off counter determination - cutting in line .]
	{\includegraphics[angle=0,width=0.5\textwidth]{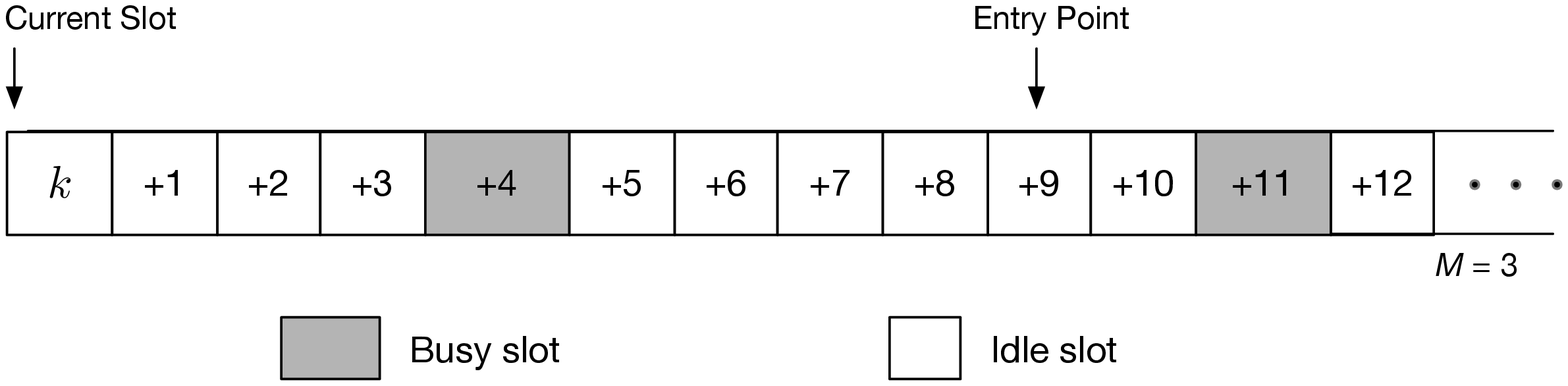}
		\label{f:EntranceP2}}
	\caption{Contention intensity estimation and initial back-off counter determination in the CIDC.}\label{f:ProtocolIllus} \vspace{-5mm}
\end{figure}

The MAC-layer design of the CIDC is based on the 802.11p MAC and inherits the slot based structure with carrier sensing, the back-off counter, and the DIFS. Details regarding the MAC layer of the 802.11p can be found in many works (e.g., \cite{J_KHafeez2013} and \cite{J_KXu2016}) and thus are neglected here. The only and key modification in the MAC layer of the CIDC is regarding how the initial back-off counter of a packet is determined. In the CIDC, the initial back-off counter is determined based on the contention intensity (as opposed to a random selection in the 802.11p MAC) as follows:
\begin{itemize}
	\item For each new packet, its initial back-off counter is set to $M$ times the contention intensity (itself included as a contending packet when calculating the contention intensity), where the protocol parameter $M$ is a positive integer.  
\end{itemize}
The above access strategy is illustrated in Figs.~\ref{f:EntranceP1}~and~\ref{f:EntranceP2}.  In both cases, $M$ is set to 3 and $n=2$ existing packets are contending. Therefore, if a new packet arrives at this instant, its entry point is the $(n+1) M = 9$th slot from the slot $k$. \textcolor{Black}{Packets at different vehicles contending for channel access can be considered as waiting in a virtual queue.}  The new packet will either enter from the end of the virtual queue of contending packets (shown in Fig.~\ref{f:EntranceP1}) or cut in line (shown in Fig.~\ref{f:EntranceP2}). 
Assuming no more new arrivals in the next 11 slots, the new packet will be transmitted after the two existing packets in the case of Fig.~\ref{f:EntranceP1} and between the two existing packets in the case of Fig.~\ref{f:EntranceP2}. \textcolor{Black}{Note that, unlike the timeline in Fig.~\ref{f:MeasureCK} which is shown from the perspective of one vehicle, Figs.~\ref{f:EntranceP1}~and~\ref{f:EntranceP2} are shown from the perspective of a virtual observer which is assumed to have real-time information of the entire vehiclular network.  For each vehicle, it has, through estimation, the information on the number of vehicles communicating with it and the instantaneous contention intensity, but not the value of the back-off counters of contending packets at other vehicles. Therefore, the entry points in Figs.~\ref{f:EntranceP1}~and~\ref{f:EntranceP2} for a new packet can be determined by the corresponding vehicle while the information on which slots are busy is not available to any vehicle but only the virtual network observer.}


Similar to the 802.11p MAC, the back-off counter of any packet reduces by one after each idle slot, freezes whenever the channel is detected busy, and unfreezes after the channel becomes idle for the length of a DIFS. The packet starts transmitting when its back-off counter reaches zero.

%

It can be seen that the proposed design is fully distributed and requires minimum modification from the 802.11p MAC. Moreover, the overhead of the proposed design in either communication or computation is very small. Specifically, the communication overhead is introduced as a result of adding the offset information  $\sigma_i$ to the safety message. A 24-bit section is more than enough to convey an offset information to the accuracy of a time slot $T_\mathrm{s}$ (typically $13\times 10^{-6}$s). Considering that a safety message packet is typically 200-300 Bytes in size at the physical layer \cite{J_YPark2013}, \cite{M_DJiang2006}, the communication overhead is thus merely $1\%-2\%$ of the packet size. Regarding the computation overhead, despite the fact that tracking the offset information of surrounding vehicles appears to be a non-negligible computation load, it is important to note that safety applications generally and naturally rely on real-time extraction and processing of the status information from all received messages, which typically yields complicated calculations (e.g.,  trajectory prediction). Therefore, the computation load of estimating real-time contention intensity, which is mostly just counting the number of received messages, can be reasonably considered as insignificant for the application layer compared to the computation load necessary to support safety applications.  

\textcolor{Black}{Note that the proposed design exploits information from other vehicles and implements coordination in a one-hop scope. Specifically, each vehicle extracts the offset information of other vehicles which are one-hop apart and accesses the channel based on estimating the instantaneous contention intensity within one hop. Therefore, the design is not targeted at addressing the hidden terminal problem, which involves vehicles two hops apart from each other, but to inherit the distributed nature and the low overhead of 802.11p while improving its performance. Nevertheless, it is possible to build on the proposed design to mitigate the hidden terminal problem. However, implementing information exchange and coordination within a two-hop scope will inevitably increase the overhead of the design.}


%
%
%

\subsection{CIDC: the Formulation}


In order to formulate the system model, the following definitions and denotations are introduced first. A list of symbols used in this paper is given in Table~\ref{t:Notation}.

\begin{table}[t!]
	\begin{center}
		\caption{List of Symbols}\label{t:Notation}
		{\setlength{\extrarowheight}{1.5pt}
			\begin{tabularx}{0.80\textwidth}{c|c}\hline\hline
				$\lambda$ & the frequency of safety message broadcast  \\ \hline
				$\sigma_i$ & the random offset of vehicle $i$, $\sigma_i \in (0, 1/\lambda]$  \\ \hline
				$N$ & number of vehicles within communication range \\ \hline
				$M$ & the CIDC protocol parameter \\ \hline
				
				$T_\mathrm{s}$ & the duration of an idle time slot \\ \hline
				$T_\mathrm{Tx}$ & the duration of a packet transmission \\ \hline
				$T_\mathrm{DIFS}$ & the duration of a DIFS\\ \hline
				$K$ & ($T_\mathrm{Tx} + T_\mathrm{DIFS}$)/$T_\mathrm{s}$  \\ \hline
				$k[s]$ & the $s$th mini-slot of slot $k$ \\ \hline
				$c(k)$ & the contention intensity at the beginning of slot $k$ \\ \hline
				$h(k)$ & indicator, $h(k)$ is $1 (0)$ if slot $k$ is busy (idle) \\ \hline
				$e(k[s])$ & initial back-off counter of a packet arriving in $k[s]$\\ \hline
				
				$n_\mathrm{I}(k)$ & number of packet arrivals in slot $k$\\ \hline
				$n_\mathrm{o}(k)$ & number of packet transmissions in slot $k$\\ \hline
				$b^\mathrm{max}(k)$ & the maximum back-off counter as slot $k$ begins \\ \hline
				$e_v(k[s])$ & the virtual entry point in slot $k[s]$\\ \hline
				$\upsilon(k[s])$ & the packet-to-slot ratio as mini-slot $k[s]$ ends\\ \hline
				$\upsilon_\mathrm{s}$ & the expected number of packets in a busy slot \\ \hline
				
				$N_\mathrm{sat}$ & the number of users at saturation point  \\ \hline
				$c_\mathrm{s}$ & the expected $c(k)$ in a steady state \\ \hline
				
				$P_\mathrm{col}$ & the collision probability  \\ \hline
				$P_\mathrm{col}^\mathrm{UB}$ & the upper bound of the collision probability  \\ \hline
				$P_\mathrm{ck}(0)$ & the probability that $c(k)=0$ \\ \hline
				
				$d_\mathrm{o}$ & the expected overall delay \\ \hline
				$d_\mathrm{c}$ & the expected contention delay \\ \hline
				
				\hline\hline
			\end{tabularx}
		}
	\end{center}
	\vspace{-0.2cm}
\end{table}

\textit{Slot and mini-slot}: A slot is the duration in which the back-off counter of a packet remains unchanged. A mini-slot is a time duration with the length of $T_\mathrm{s}$. If the channel is idle, a slot consists of one mini-slot. Otherwise, a slot has $K$ mini-slots. The $s$th mini-slot in slot $k$ is denoted as $k[s]$, and the set of indexes of all mini-slots in slot $k$ is denoted as $\mathcal{S}_k$. Accordingly, $\mathcal{S}_k = \{1\}$ if slot $k$ is idle, and $\mathcal{S}_k = \{1, \dots, K\}$ if slot $k$ is busy. 

\textit{Absolute and relative slot index}: The absolute slot index is an index with respect to the beginning of the protocol execution. The relative slot index is an index with respect to the current absolute index. For example, the $k$ in Figs.~\ref{f:EntranceP1}~and~\ref{f:EntranceP2} is an absolute index while the $1, 2\dots$ after the ``$+$'' are the relative slot indexes and translate to $k+1, k+2, \dots$ in absolute slot index.  

Denote the number of packets contending for channel access measured at the \emph{beginning} of slot $k$ as $c(k)$. Let $h(k) = 1$ and $h(k) = 0$ represent the events that slot $k$ is busy and idle, respectively. \textcolor{Black}{The initial back-off counter of a packet arriving at mini-slot $k[s]$, which depends on the instantaneous contention intensity, is denoted as $e(k[s])$ and referred to as the entry point of this packet.} Denote the number of packets arrived in a mini-slot $k[s]$ measured at the \emph{end} of the mini-slot and the number of packets arrived in slot $k$ measured at the \emph{end} of the slot as $n_\mathrm{I}(k[s])$ and $n_\mathrm{I}(k)$, respectively. Denote the number of packets with their back-off counter reduced to 0 in slot $k$ as $n_\mathrm{o}(k)$. Note that $h(k) = 1$ if and only if $n_\mathrm{o}(k) > 0$. The system implementing the CIDC is governed by the following rules:
\begin{subequations}\label{e:sysEqs}
	\begin{align}
	& c(k + 1)  = c(k) + \sum\limits_{l \in \mathcal{S}_k} n_\mathrm{I}(k[l]) - n_\mathrm{o}(k)   \label{e:sysE1} \\
	& e(k[s]) = M \bigg( c(k) + \!\! \sum\limits_{l =1, l \in \mathcal{S}_k}^{s} \! n_\mathrm{I}(k[l])  \bigg), \; \text{if}\;\;  n_\mathrm{I}(k[s]) \!>\! 0  \label{e:sysE2}   \\ 
	& h(k + e(k[s])) = 1, \quad \text{if}\quad  n_\mathrm{I}(k[s]) > 0 \label{e:sysE4} \\
	& n_\mathrm{o}(k \!+\! e(k[s])) \leftarrow n_\mathrm{o}(k \!+\! e(k[s])) \!+\! 1, \;\; \text{if}\;\; n_\mathrm{I}(k[s]) > 0  \label{e:sysE5}
	\end{align}
\end{subequations}
where $\leftarrow$ denotes the operation that assigns the value of the expression on the right-hand side to the variable on the left-hand side.

Equation \eqref{e:sysE1} characterizes the change in the contention intensity after a slot. Equation \eqref{e:sysE2} characterizes the proposed initial back-off counter selection rule based on the contention intensity. Equations~\eqref{e:sysE4}~and \eqref{e:sysE5} reflect the consequence of the back-off counter selection on the system status at a future time instant, i.e., the channel will be busy after $e(k[s])$ slots and one more packet will be sent in the corresponding slot. Note that $h(k)$ and $n_\mathrm{o}(k)$ should be set to 0 for all $k$ at initialization. In addition, $e(k[s])$ is only defined for the mini-slots with packet arrivals. The event that more than one packet arrives in a mini-slot (i.e., $13\mu$s) is neglected as the probability of such an event is extremely small. Correspondingly,  $n_\mathrm{I}(k[s])$ is equal to either zero or one.

Another variable of interest is the maximum back-off counter among all contending packets measured at the \emph{beginning} of slot $k$, denoted as $b^{\max}(k)$.  The maximum back-off counter is governed by the following rule:
\begin{align}
 b^{\max}(k \! + \! 1) \! = \!\! \left\{
\begin{array}{ll}
\!\!\!\max\bigg\{\!b^{\max}(k) \!-\! 1, 0\bigg\} , \;\text{if}\; \sum\limits_{l \in \mathcal{S}_k} n_\mathrm{I}(k[l]) = 0 \\
\!\!\!\max\bigg\{\! \max\limits_{l \in \mathcal{S}_k}\{e(k[l])\}, b^{\max}(k)\!\bigg\} \!-\! 1, \;\,\text{else}  \\
\end{array}
\right. \label{e:sysE3}
\end{align}

From the equations \eqref{e:sysEqs} and \eqref{e:sysE3}, it can be seen that the initial back-off counter selection based on the contention intensity in the CIDC has a significant impact on the system model. First, as shown by equations \eqref{e:sysE1} and \eqref{e:sysE2}, the initial back-off counter of a packet arriving at slot $k$ is no longer random and independently decided. Instead, it is deterministic and decided by the system states through $c(k)$ and $n_\mathrm{I}(k[l]), \forall l \in \mathcal{S}_k$. This introduces a strong coupling among the back-off counters of contending packets at different vehicles. Second, as shown by equations \eqref{e:sysE4} and \eqref{e:sysE5}, the impact of the proposed initial back-off counter selection extends, again deterministically rather than randomly, to a future slot. Therefore, the system state at any given instant is dependent on the system states and events at many previous instants (possibly infinitely many previous instants depending on the channel load). 

The above coupling among the back-off counters at different vehicles and among the events across the time domain cannot be characterized by the classic packet-perspective Markov chain based modeling. Next, we will analyze the system performance from a network-perspective based on the concepts of packet-to-slot ratio and contention intensity.


\section{Performance Analysis}\label{s:perform}


\subsection{Basic Features}

The proposed CIDC has two intuitive features.

\begin{figure*}[!t]
	\centering \subfloat[Collision can only happen at slot $k+xM$, where $x \in \{2, 3, \dots \}$.]
	{\includegraphics[angle=0,width=0.56\textwidth]{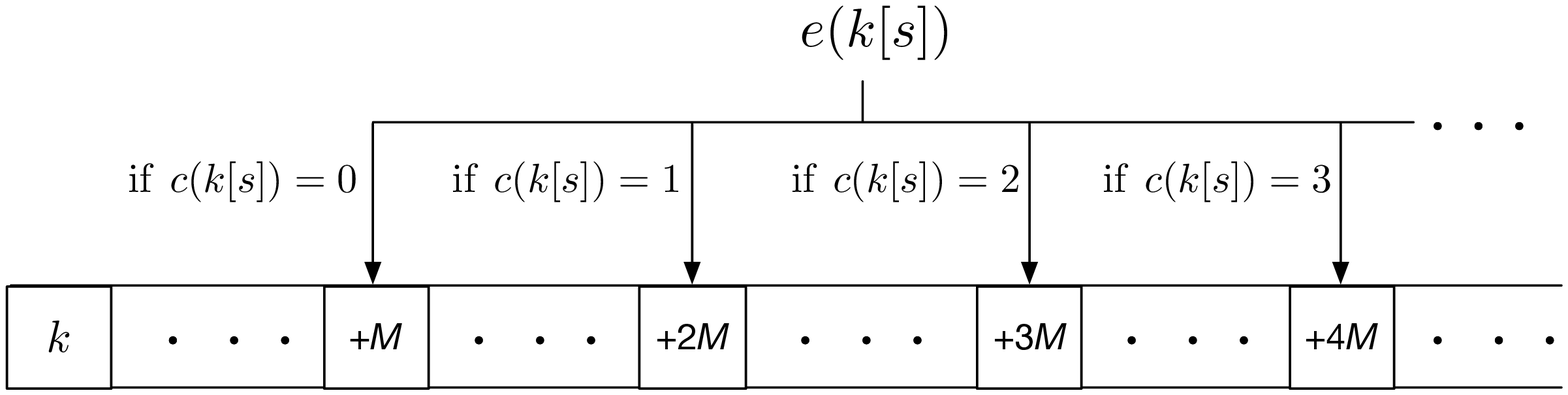}
		\label{f:CollisionLocations}}
	\hspace{1mm} 
	\subfloat[Collision can happen when $c(k)$ has been reducing. ]
	{\includegraphics[angle=0,width=0.56\textwidth]{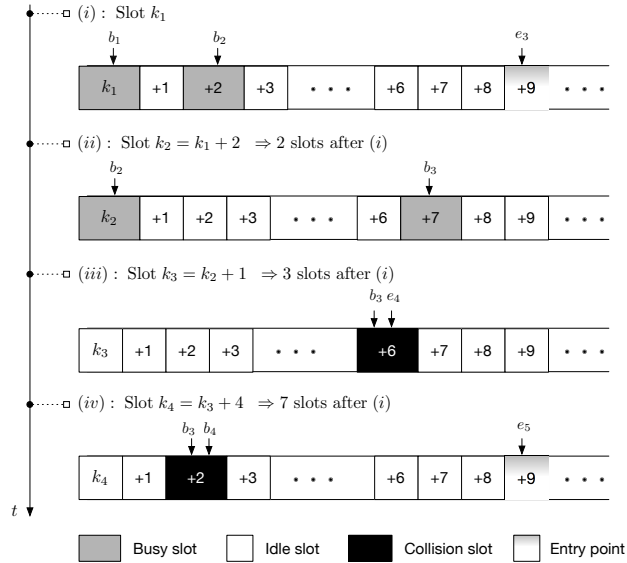}
		\label{f:CollisionCond}}
	\vspace{1mm}
	\caption{Illustration of the two basic features of the CIDC.}\label{f:featureIllus} \vspace{-5mm}
\end{figure*}

First, a packet collision can only happen at certain slots. Specifically, since a packet arriving at mini-slot $k[s]$ sets its initial back-off counter based on $e(k[s])$ in \eqref{e:sysE2}, a collision with existing packets can only happen at slots $k + 2M, k + 3M, \dots$. This is illustrated in Fig.~\ref{f:CollisionLocations}, where $c(k[s]) = e(k[s])/M$.


Second, a packet collision can only happen when there has been a decrease in the contention intensity $c(k)$. If the number of packets contending for channel access is constant or steadily increasing, a collision cannot happen. This is illustrated in Fig.~\ref{f:CollisionCond}. In this figure, $M=3$, $b_i$ represents the back-off counter of the $i$th contending packet at the current instant, and $e_i$ represents the initial back-off counter of the $i$th contending packet arriving at the current slot. At Step ($i$), as there are two existing packets with remaining back-off counters at $b_1 =0$ and $b_2 = 2$, respectively, the initial back-off counter of the arriving new packet, denoted by $e_3$, is set to 9. At Step ($ii$), i.e., two slots after Step ($i$), the packet corresponding to the back-off counter $b_1$ has been transmitted, and the packet with the back-off counter $b_2$ is transmitting as $b_2$ has reduced to 0. The remaining back-off counter of the packet with the initial back-off counter $e_3$, denoted by $b_3$, is at 7. After another slot, i.e., at Step ($iii$), a new packet arrives and sets its initial back-off counter $e_4$ to 6 since there is only one existing packet contending for the channel access. Unfortunately, the back-off counter of the existing packet, i.e., $b_3$, also decreases to 6 at this slot. This causes a collision that will happen in 6 slots. At Step ($iv$), another packet arrives and sets its initial back-off counter $e_5$ to 9. It can be observed that the contention intensity has been decreasing from Steps~($i$) till before Step~($iii$), which renders the packet collision possible. By contrast, the contention intensity increases between Step~($iii$) and Step~($iv$), and thus the new packet arriving at Step~($iv$) does not encounter a collision.


Next, we will characterize the system given by \eqref{e:sysEqs} in detail. 

%

\subsection{Packet-to-Slot Ratio}

As a network-perspective parameter, the packet-to-slot ratio is introduced here to represent the number of simultaneously contending packets over the number of slots that accommodate the back-off counters of these contending packets. In the case of the 802.11p MAC, the packet-to-slot ratio can be simply defined as the number of packets in the fixed range $[0, W- 1]$ divided by $W$, where $W$ is the contention window. In the CIDC, there is no contention window and both the number of packets and the range of their back-off counters vary over time.

In order to characterize the range of the back-off counters, we introduce the virtual entry point $e_v(k[s])$ for any mini-slot defined as:
\begin{align}
e_v(k[s]) = \left\{
\begin{array}{ll}
\!\! M \bigg( c(k) + 1 \bigg),  \quad \text{if}\quad h(k) = 0 \\
\!\! M \bigg( c(k) + \sum\limits_{l= 1, l \in \mathcal{S}_k}^{s - 1} n_\mathrm{I}(k[l]) + 1 \bigg),  \;\; \text{else}  \\
\end{array}
\right.
\end{align}
Note that $e_v(k[s])$ is the entry point \emph{if} a packet were to arrive at mini-slot $k_s$ and thus is defined for any $k[s]$ (By contrast, $e(k[s])$ is only defined for a $k[s]$ if $n_\mathrm{I}(k[s]) > 0$).

The packet-to-slot ratio, measured at the \textit{end} of a mini-slot $k[s]$, is then defined as follows:
\begin{align}\label{e:upsilonDef}
\upsilon(k[s]) = \frac{c(k) + \sum\limits_{l = 1, l \in \mathcal{S}_k}^{s} n_\mathrm{I}(k[l])}{\max\{b^{\max}(k), e_v(k[s])\}}.
\end{align}

The above definition characterizes the average number of packets per slot, which depends on $N$ and $\lambda$. While the protocol parameter $M$ is also expected to have an impact on the packet-to-slot ratio, such impact is not straightforward based on the equation \eqref{e:upsilonDef}. The following lemma shows that the impact of $M$ becomes evident as the system approaches saturation.

\textbf{Lemma~1}: The packet-to-slot ratio $\upsilon(k[s])$ is bounded by $1/M$ in the CIDC.

\emph{Proof}: See Section~\ref{s:ProofLemma1} in Appendix.

Lemma~1 shows that the packet-to-slot ratio $\upsilon(k[s])$ saturates at $1/M$ for the CIDC. When the channel load is small, $\upsilon(k[s])$ is determined by the number of vehicles $N$, message frequency $\lambda$, etc. while the parameter $M$ does not have a significant impact. As the channel load increases, the impact of $M$ on the packet-to-slot ratio becomes more significant. This feature has a significant impact on the collision probability of the CIDC and will be discussed in detail in Subsection~\ref{ss:ColCondP}.

It can be shown that the $1/M$ busy ratio is a bound that cannot be achieved. Given $M$, $N$, and $\lambda$, the expected value of $\upsilon(k[s])$ in a steady state \footnote{The term steady state refers to the state when 1) the system is not over-saturated (an increasing number of packets at the vehicles are expired by subsequent new packets in an over-saturated system); and 2) the protocol has been executed for a sufficiently long time with the fixed $N$, $\lambda$, and $M$.}, denoted as $\upsilon_\mathrm{s}$, is given by the following lemma.

\textbf{Lemma~2}: Given $M$, $N$, and $\lambda$, the expected packet-to-slot ratio in a steady state is given by:
\begin{align}\label{e:upsilonSteady}
\upsilon_\mathrm{s} = \frac{1}{1- P_\mathrm{ck}(0)} \frac{\lambda N T_\mathrm{s}}{n_\mathrm{s} - \lambda N (K-1)T_\mathrm{s}}.
\end{align}
where $P_\mathrm{ck}(0)$ denotes the probability of $c(k)$ being 0 and $n_\mathrm{s}$ is the expected number of packets in a busy slot, i.e., the expected value of $n_\mathrm{o}(k)$ in the equation \eqref{e:sysE1}. 

\emph{Proof}: See Section~\ref{s:ProofLemma2} in Appendix.

Note that the impact of $M$ on $\upsilon_\mathrm{s}$ is manifested through $n_\mathrm{s}$. Based on Lemma~2, the number of vehicles $N$ that leads to saturation satisfies
\begin{align}
N_\mathrm{sat} = \frac{n_\mathrm{s}}{\lambda( \frac{M}{1- P_\mathrm{ck}(0)} + K -1)T_\mathrm{s}}.
\end{align}
At saturation, as $P_\mathrm{ck}(0)$ becomes very small, an approximation of the above can be obtained by setting $P_\mathrm{ck}(0) =0$.


The expected number of packets per busy slot, i.e., $n_\mathrm{s}$, is greater than 1 due to the probability of packet collision. Specifically, it is connected to the probability that $l$ packets collide in one slot, denoted as $P^l_\text{col}$, through the following equation
\begin{align}
n_\mathrm{s} =  1 + \sum\limits_{l} (l-1) P^l_\mathrm{col}.
\end{align}
Neglecting the cases in which more than two packets collide, $P_\mathrm{col}$ becomes $P^2_\mathrm{col}$ and the above equation reduces to
\begin{align}\label{e:ns2}
n_\mathrm{s} =  1 + P_\mathrm{col}.
\end{align}

\subsection{Contention Intensity and Contention Delay}

While the packet-to-slot ratio $\upsilon(k[s])$ can be an indicator of the channel load, a full characterization of the system performance requires additional metrics. For example, the contention delay of a packet depends on the exact number of packets contending for channel access, i.e., the contention intensity $c(k)$. 

The contention intensity can be characterized by the one-slot probability transition matrix of $c(k)$. Specifically, given $c(k)$, the following cases of $c(k+1)$ are possible:
\begin{itemize}
	\item [i] the $k$th slot is busy and there is $n_\alpha(k)$ arrivals in the slot $k$, then $c(k+1) = c(k) + n_\alpha(k) - n_\mathrm{o}(k)$
	\item [i] the $k$th slot is busy and there is no arrival in the slot $k$, then $c(k+1) = c(k)  -  n_\mathrm{o}(k)$
	\item [iii] the $k$th slot is idle and there is $n_\beta(k)$ arrivals in the slot $k$, $c(k+1) = c(k) + n_\beta(k)$
	\item [iv] the $k$th slot is idle and there is no arrival in the slot $k$, $c(k+1) = c(k)$.
\end{itemize}

Consider the event that $c(k) = j$ packets in slot $k$ and $c(k+1) = i$ packets in slot $k + 1$. Denote the probabilities of such an event given that slot $k$ is idle and busy as $P^\mathrm{I}_{k+1, k}(i, j)$ and $P^\mathrm{B}_{k+1, k}(i, j)$, respectively. Neglect the probability of more than 1 packet arriving in the same mini-slot. Then, $P^\mathrm{I}_{k+1, k}(i, j)$ and $P^\mathrm{B}_{k+1, k}(i, j)$ with $i, j \in \{1, \dots, N\}$ are given by
\begin{align}
P^\mathrm{I}_{k+1, k}(i, j) =
\left\{
\begin{array}{ll}
\! 0 , \quad\; \text{if}\quad  i < j  \quad \\
\! P^\mathrm{I}(i -j), \quad \text{else}
\end{array}
\right.
\end{align}
and
\begin{align}
&P^\mathrm{B}_{k+1, k}(i, j) \!  \nonumber \\
&= \!\left\{
\begin{array}{ll}
\!\! 0, \qquad\qquad \text{if} \quad i < j - 2 \\
\!\! P_\mathrm{col} P^\mathrm{B}(0), \qquad \text{if} \quad i = j - 2 \\
\!\! P_\mathrm{col}P^\mathrm{B}(i \!-\! j \!+\! 2) + (1 \!-\! P_\mathrm{col})P^\mathrm{B}(i \!-\! j \!+\! 1), \; \text{else}
\end{array}
\right.
\end{align}
where
\begin{subequations}\label{e:PIPBdef}
\begin{align}
 P^\mathrm{I}(x) &= \binom{N}{x} (\lambda T_\mathrm{s})^{x} (1-\lambda T_\mathrm{s})^{N - x}, \\
\!\!  P^\mathrm{B}(x) &= \binom{N}{x} (\lambda K T_\mathrm{s})^{x} (1-\lambda K T_\mathrm{s})^{N - x}.
\end{align}
\end{subequations}
Since a slot is busy with probability $\upsilon_\mathrm{s}$, the overall one-slot probability transition matrix of $c(k)$ is given by
\begin{align}\label{e:ckProbTrans}
\mathbf{P}_{k+1, k} = \bigg[ (1 - \upsilon_\mathrm{s})P^\mathrm{I}_{k+1, k}(i, j) + \upsilon_\mathrm{s}P^\mathrm{B}_{k+1, k}(i, j) \bigg]_{i, j}
\end{align}

Denote $\mathbf{p}^\mathbf{c}_\mathrm{s} = [P(c(0)), \dots, P(c(N))]^\text{T}$ as the vector of the steady-state probability of $c(k), k = 0, \dots, N$. Then $\mathbf{p}^\mathbf{c}_\mathrm{s}$ satisfies
\begin{align}\label{e:ckInstant}
(\mathbf{P}_{k+1, k} - \mathbf{I}) \mathbf{p}^\mathbf{c}_\mathrm{s} = 0
\end{align}
Therefore,  $\mathbf{p}^\mathbf{c}_\mathrm{s}$ is in the null space of $\mathbf{P}_{k+1, k} - \mathbf{I}$. 

The expected contention intensity over all slots, i.e., the expected  $c(k)$ denoted as $c_\mathrm{s}$, and its relation with the average packet delay given $N$, $\lambda$, and $M$ are given by the following theorem.

\textbf{Theorem~1}: The average contention intensity $c_\mathrm{s}$ and the average overall packet delay $d_\mathrm{o}$ of the CIDC can be solved from the following equations:
\begin{subequations}
\begin{align}
&d_\mathrm{o} \!=\! \bigg(\!c_\mathrm{s} \!+\! 1 \!-\! \frac{1 \!-\!  P_\mathrm{c_k}(0)}{2}\!\bigg) K T_\mathrm{s}  \!+\! \bigg(\! M(c_\mathrm{s} \!+\! 1) \!-\! c_\mathrm{s} \! \bigg) T_\mathrm{s}   \label{e:delayOverall}\\
&\qquad\qquad\qquad  N\lambda d_\mathrm{o} = c_\mathrm{s}     \label{e:delayCs} \\
&\qquad \qquad\qquad P_\mathrm{c_k}(0) = \left(1 - \frac{c_\mathrm{s}}{N}\right)^N.   \label{e:delayPck0} 
\end{align} 
\end{subequations}

\emph{Proof}: See Section~\ref{s:ProofTheorem1} in Appendix.

The overall packet delay $d_\mathrm{o}$ is the time duration between the instant that a packet arrives and the instant that the packet transmission completes. The contention delay $d_\mathrm{c}$ is the time duration between the instant that a packet arrives and the instant that the packet transmission begins. The relation between the two delay metrics is given by:
\begin{align}
d_\mathrm{c} = d_\mathrm{o} - K T_\mathrm{s} +  T_\mathrm{DIFS}.
\end{align}   

Theorem~1 characterizes the relation between the expected overall delay $d_\mathrm{o}$, the expected number of contending packets $c_\mathrm{s}$, and the probability of no packet contending for channel access  $P_\mathrm{c_k}(0)$. In order to obtain further insight, approximations for $c_\mathrm{s}$ in the cases of a small $N$ and a large $N$ can be derived, respectively. 

\textbf{Lemma~3}: The small-$N$ and large-$N$ approximations of the expect number of contending packets $c_\mathrm{s}$, denoted as $c^\text{L}_\mathrm{s}$ and $c^\text{H}_\mathrm{s}$, respectively, can be given as follows:
\begin{subequations}
\begin{align}
c^\text{L}_\mathrm{s} = \frac{N\lambda (K+M) T_\mathrm{s}}{ 1 - N\lambda (K+M - 1) T_\mathrm{s}}  \label{e:smallNcs}\\
c^\text{H}_\mathrm{s} = \frac{N\lambda (K/2+M) T_\mathrm{s}}{ 1 - N\lambda (K+M - 1) T_\mathrm{s}}  \label{e:largeNcs}
\end{align}
\end{subequations}

\emph{Proof}: See Section~\ref{s:ProofLemma3} in Appendix.

The delay in 802.11p can be obtained similarly for a comparison. In fact, the overall delay in 802.11p can be solved from \eqref{e:delayOverall}-\eqref{e:delayPck0} by substituting  $M(c_\mathrm{s} + 1)$ with $W/2$ where $W$ is the contention window. Therefore, with a proper choice of $M$, the CIDC should have a smaller delay than that in 802.11p when $c_\mathrm{s}$, or ultimately $N$ and $\lambda$, is not too large. This will be verified in Section~\ref{s:simu}~Simulations. 

It should be noted that Theorem~1 holds under the assumption that the collision probability is not so high that it has a significant impact on the average delay. If the system is beyond saturation, Theorem~1 does not apply.

%

\subsection{Collision Conditions and Probability}\label{ss:ColCondP}

In this section, the instantaneous collision probability is analyzed first. Then, the upper bound on the collision probability is derived in a closed form. 

Consider the event that packet $B$, which arrives at the mini-slot $k_2[s_2]$, collides with packet $A$, which arrives at mini-slot $k_1[s_1]$, where $k_1 < k_2$. Let $\alpha = k_2 - k_1$. Suppose there are $\beta$ busy slots in the interval $[k_1[s_1], k_2[s_2])$, in which $\tau$ packets are transmitted. Denote the number of packet arrival in the interval $[k_1[s_1], k_2[s_2]]$ as $\eta$.

\textbf{Lemma~4}: Packets $A$ and $B$ can collide if and only if
\begin{align}\label{e:BetaAlphaM}
M (\tau - \eta)= \alpha.
\end{align}

\emph{Proof}: See Section~\ref{s:ProofLemma4} in Appendix.

Based on Lemma~4, if the gap between the arrival slot index of the two packets is not a multiple of $M$, the two packets have no chance to collide. Moreover, whether a collision can happen or not in the CIDC depends on both the time instant of the arrival and the number of recent transmission and arrival events, as suggested by \eqref{e:BetaAlphaM}. The result is that collision can happen only at specific slots and under specific conditions.

Lemma~4 verifies the aforementioned feature that a duration with more transmissions than arrivals is necessary for a collision to happen (as shown in Fig.~\ref{f:CollisionCond}). Moreover, as the gap $\alpha$ increases, a larger difference between the number of transmission and arrival events, i.e., a larger $\tau - \eta$, is required for the collision to happen according to \eqref{e:BetaAlphaM}. Considering the fact that the average packet transmission rate and packet arrival rate should be equal as long as the system is not beyond saturation, a smaller collision probability is implied for packets with a larger gap between their arrival slot indexes. This suggests that the CIDC effectively limits the \textit{collision range}. By contrast, any two contending packets can collide in the case of 802.11p MAC. 

The above insight reveals the difference in the underlying causes of a packet collision in the 802.11p and the CIDC. In the 802.11p, a packet collision is caused by accommodating contending packets in a fixed contention window. Consequently, the \textit{number} of vehicles $N$ has a direct and definitive impact on the collision probability. \textcolor{Black}{By contrast, the cause of a packet collision in the CIDC is the \emph{uneven} intervals between successive packet arrival instants in the random packet arrival process. This unevenness becomes a limiting factor in the coordination of packet transmissions in a distributed approach. The impact of the number of vehicles is not definitive. For instance, a collision will not happen, regardless of $N$, if the number of contending packets steadily increases. However, a larger $N$ generally increases the probability that condition \eqref{e:BetaAlphaM} is satisfied and therefore leads to a larger collision probability.}

A packet can be involved in a collision in two scenarios. The first scenario is when the packet collides with an existing contending packet upon its arrival, referred to as forward packet collision (e.g., the packet corresponding to $e_4$ in Step~$(iii)$ of Fig.~\ref{f:CollisionCond}). The second scenario is when the packet is involved in a collision with a packet arriving after it during its back-off, referred to as a backward packet collision (e.g., the packet corresponding to $b_3$ in Step~$(iii)$ of Fig.~\ref{f:CollisionCond}). Since a forward packet collision for one packet must be a backward collision for another, the probability of either type of collision is identical. 


Consider a forward collision of packet $B$ with packet $A$ in the example preceding Lemma~4 but without assuming packet $B$'s arrival at a specific slot. Define $c_1 = c(k_1) + \sum_{l \leq s_1}  n_\mathrm{I}(k_1[l])$. The set of $\alpha$ so that packet $B$ can forward-collide with packet $A$ is $\{M, 2M, \dots, (c_1-2)M \}$. \textcolor{Black}{If packet $B$ arrives less than $M$ slots after packet $A$ arrives, the condition \eqref{e:BetaAlphaM} cannot be satisfied. If packet $B$ arrives more than $(c_1-2)M$ slots after packet $A$ arrives, the back-off counter of packet $A$  reduces to less than $2M$ and collision cannot happen either (submitting $\alpha > (c_1-2)M$ into \eqref{e:BetaAlphaM} leads to the result $\tau - \eta > c_1 -1$, which cannot be satisfied.).} 
For a given $\alpha$ and a given $c_1$, a forward collision happens when: 1) packet $B$ arrives; and 2) the condition \eqref{e:BetaAlphaM} is satisfied. Denote the probability that $i$ packets are transmitted and $j$ packets arrive in the $\alpha$ slots as $P^{\alpha}_\mathrm{T, A}(i, j)$. The forward-collision probability given $\alpha$ and $c_1$ can be written as follows:
\begin{align}\label{e:PcolAlpha}
P_{\text{col}}^{\alpha,c_1}(\tau,  \eta) = P_\mathrm{A}^{k_2}  \sum\limits_{\tau = \frac{\alpha}{M}}^{c_1^\prime} \!  P^{\alpha}_\mathrm{T, A}(\tau, \tau \!-\! \frac{\alpha}{M}),
\end{align}
where $c_1^\prime = c_1 -2$ and $P_\mathrm{A}^{k_2}$ represents the probability of at least one arrival event occurring in slot $k_2$. 

In equation \eqref{e:PcolAlpha}, $P_\mathrm{A}^{k_2}$ depends on whether slot $k_2$ is busy or not. The probability $P^{\alpha}_\mathrm{T, A}(\tau, \tau -\alpha/M)$ depends on two random processes, i.e., the packet arrival and transmission processes. The transmission process is dependent on the arrival process and can be considered as the output of the arrival process after going through the contention mechanism. The output of an random process going through a general nonlinear system can be intractable. However, despite the dependence between the two processes, the following two properties can be used to derive the probability in \eqref{e:PcolAlpha}. First, except for rare cases, the number of transmissions in the $\alpha$ slots is decided by the number of arrivals in a non-overlapping duration. Due to the independent increment property of the random arrival process, the number of transmissions and arrivals in the $\alpha$ slots can be considered as independent, i.e.,
\begin{align}\label{e:PcolTAindepend}
P^{\alpha}_\mathrm{T, A}(\tau, \tau \!-\! \frac{\alpha}{M}) = P^{\alpha}_\mathrm{T}(\tau) P^{\alpha}_\mathrm{A}(\tau \!-\! \frac{\alpha}{M})
\end{align}
where $P^{\alpha}_\mathrm{T}(i)$ and $P^{\alpha}_\mathrm{A}(j)$ denote the probabilities that $i$ packets are transmitted and that $j$ packets arrive in the $\alpha$ slots, respectively. Second, the average packet transmission rate must be equal to $N\lambda$ unless the system is in a state beyond saturation. Based on the above, the probability $P^{\alpha}_\mathrm{T}(\tau)$ can be obtained based on an appropriate approximated process.

Define $\beta = \alpha/M$. The overall collision probability is
\begin{align}\label{e:PcolIns}
P_\text{col}  = 2\sum\limits_{c_1^\prime = 1}^{N} P(c_1^\prime) \sum\limits_{\beta = 1}^{c_1^\prime} P_{\text{col}}^{\alpha, c_1}(\tau, \eta),
\end{align}
where $P(c_1^\prime)$ can be found from the steady state probability distribution of $c(k)$, i.e., $\mathbf{p}^\mathbf{c}_\mathrm{s}$ in \eqref{e:ckInstant}. In \eqref{e:PcolIns}, the summation over $c_1^\prime$ starting from 1 is equivalent to a summation over $c_1$ starting from 3 as it can be shown that no packet can forward-collide with packet $A$ if the initial back-off counter of packet $A$ is $M$ or $2M$, corresponding to $c_1=1$ and $c_1 = 2$, respectively.  

Note that the equations \eqref{e:upsilonSteady}, \eqref{e:ns2}, \eqref{e:ckInstant}, and \eqref{e:PcolIns} form a complete set of equations to find a solution for $\upsilon_\mathrm{s}$, $n_\mathrm{s}$, $P_\text{col}$, and $\mathbf{p}^\mathbf{c}_\mathrm{s}$. A solution can be found numerically. However, as a numeric solution only provides limited insight, the following upper bound on the average collision probability in the CIDC is derived.

\textbf{Theorem~2}: The average collision probability of the CIDC, assuming that the system is not beyond saturation, is upper-bounded by
\begin{align}\label{e:PcolUB}
P_\text{col}^\text{UB} = & \bigg( \frac{(a_1 + 1 + b_{K-1})^2}{4} + \frac{b_1 (a_K - a_1)}{ 1 - P_\mathrm{c_k}(0)}  \nonumber\\ 
& - (a_1 + 1) b_{K-1} \bigg)^\frac{1}{2} + \frac{a_1 + 1 + b_{K-1}}{2} - 1
\end{align}
where $P_\mathrm{c_k}(0)$ can be found from equations \eqref{e:delayOverall}-\eqref{e:delayPck0}, $b_1 = \lambda N T_\mathrm{s}$, $b_{K-1}=  \lambda N (K-1)T_\mathrm{s}$, and
\begin{subequations}
\begin{align}
a_ 1 &=  (1 - P_\mathrm{c_k}(0))    (1 - (1 - \lambda T_\mathrm{s})^N ) \\
a_K &= (1 - P_\mathrm{c_k}(0)) (1 - (1 - \lambda K T_\mathrm{s})^N). 
\end{align}
\end{subequations}

\emph{Proof}: See Section~\ref{s:ProofTheorem2} in Appendix.

\section{Simulations}\label{s:simu}

In this section, the performance of the CIDC is demonstrated and compared with both the analytical results and the performance of the 802.11p. First, the collision probability and contention delay are simulated assuming an accurate estimation of contention intensity. Then, the analytical results are compared to the numerical results. At last, the performance is simulated with errors in the estimated contention intensity caused by factors such as the vehicle mobility. 

The following general parameter setting is used: the safety message broadcast frequency $\lambda$ is $10$Hz; the length of a DIFS $T_\mathrm{DIFS}$ is $58\mu$s; and the length of a time slot $T_\mathrm{s}$ is $13\mu$s. Two packet transmission durations are considered: $T_\mathrm{Tx} = 254 \mu$s ($K = 24$) and $T_\mathrm{Tx} = 332 \mu$s ($K = 30$). Assuming a transmission data rate of 6Mb/s, these two transmission durations correspond to a physical-layer packet length of 190 Bytes and 250 Bytes, respectively. The protocol parameter $M$ is set to 2 in all simulations.  

\subsection{Performance with Accurate Contention Intensity Estimation}

The first example demonstrates the collision and delay performance of the CIDC and the comparison with the 802.11p MAC. The results are averaged over 10 simulation rounds with 160 message cycles used in each round. The number of vehicles varies from 25 to 250. 

\begin{figure}[!t]
	\centering \subfloat[Collision probability versus $N$ averaged over 10 rounds, $T_\mathrm{Tx} = 254 \mu$s ($K = 24$).]
	{\includegraphics[angle=0,width=0.43\textwidth]{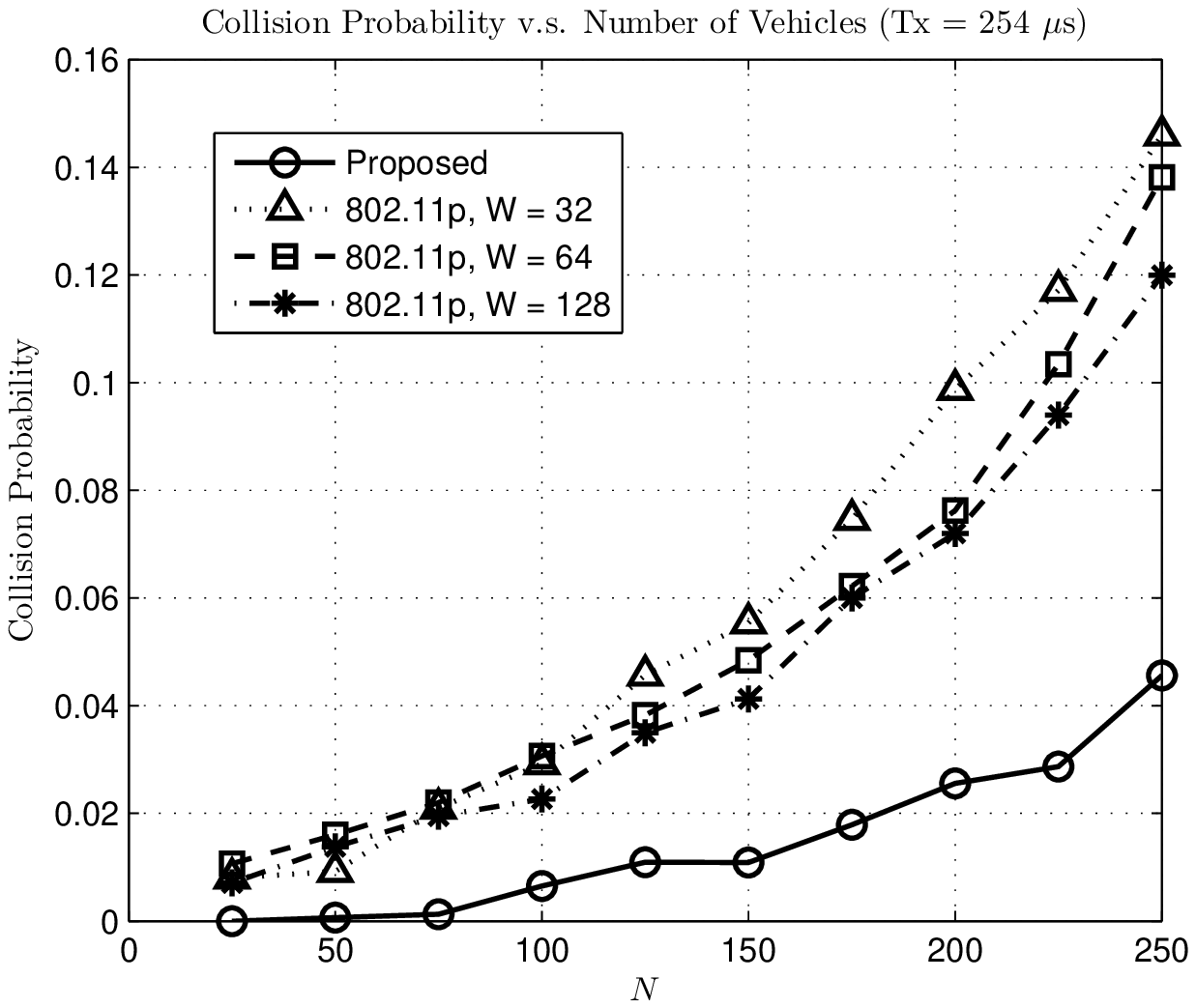}
		\label{f:Pcol254}}
	\hspace{1mm} 
	\subfloat[Collision probability versus $N$ averaged over 10 rounds, $T_\mathrm{Tx} = 332 \mu$s ($K = 30$).]
	{\includegraphics[angle=0,width=0.43\textwidth]{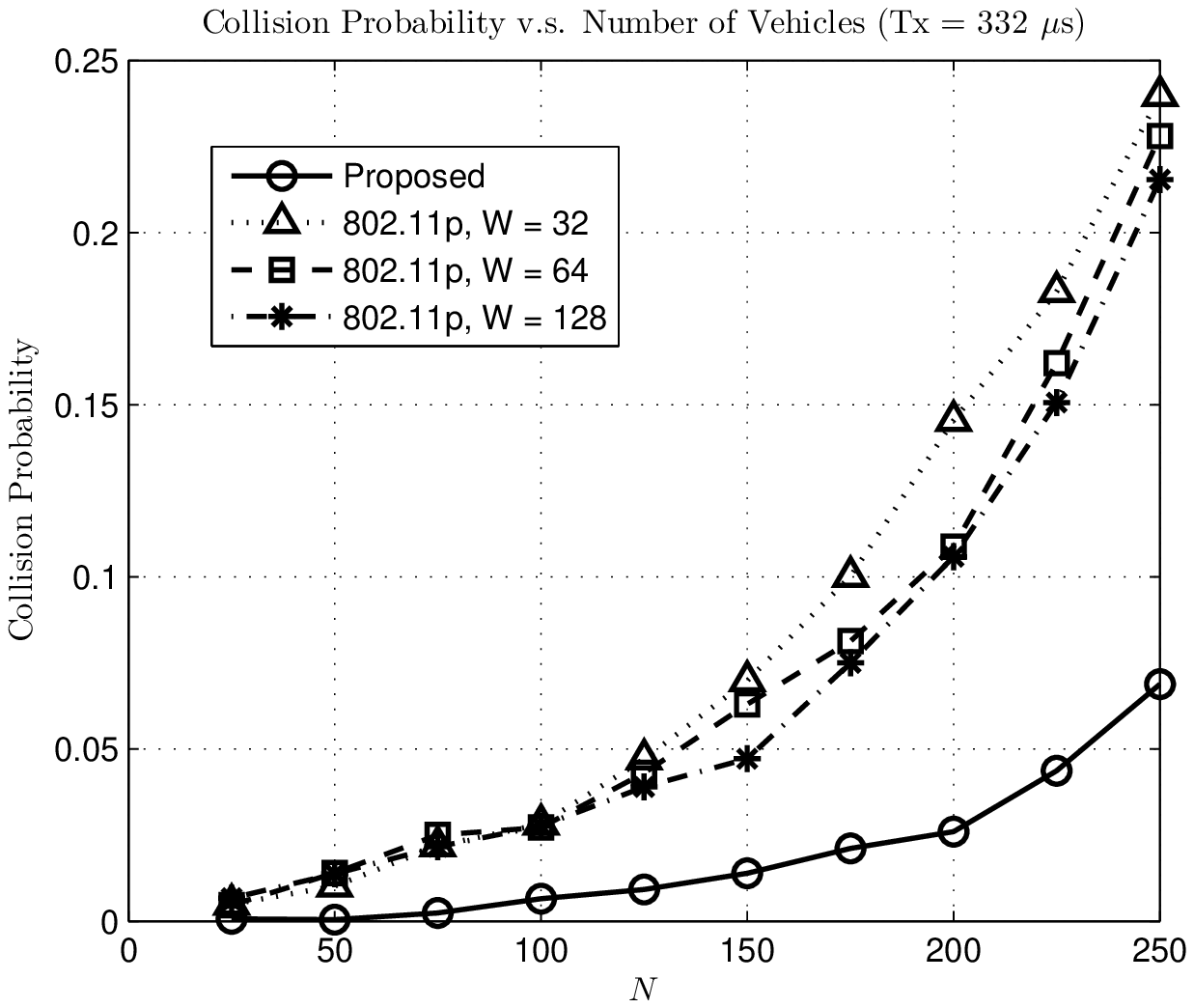}
		\label{f:Pcol332}}
	\vspace{1mm}
	\caption{Collision probability versus number of vehicles under two different packet transmission time durations.}\label{f:Pcol} \vspace{-5mm}
\end{figure}

Figs.~\ref{f:Pcol254}~and~\ref{f:Pcol332} demonstrate the average collision probability versus $N$ for $T_\mathrm{Tx} = 254 \mu$s ($K = 24$) and $T_\mathrm{Tx} = 332 \mu$s ($K = 30$), respectively. In each plot, the collision probability of the CIDC is compared with that of 802.11p MAC with contention windows $W=32, W=64$, and $W=128$, respectively. From the two figures, it can be seen that the CIDC has a substantially lower collision probability than that of 802.11p, regardless of the chosen contention window $W$, especially when $N$ is large.

\begin{figure}[!t]
	\centering \subfloat[Delay versus $N$ averaged over 10 rounds, $T_\mathrm{Tx} = 254 \mu$s ($K = 24$).]
	{\includegraphics[angle=0,width=0.43\textwidth]{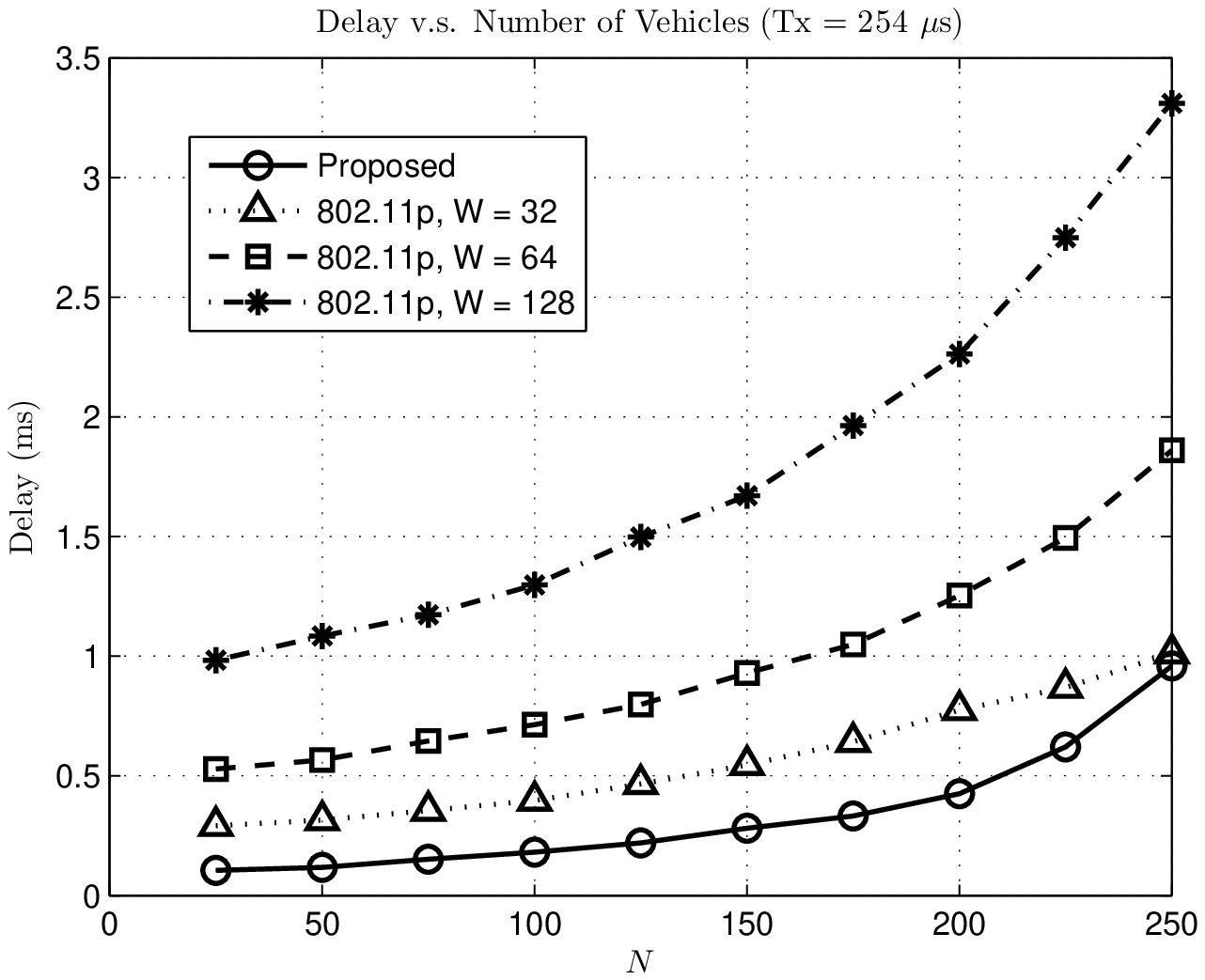}
		\label{f:Delay254}}
	\hspace{1mm} 
	\subfloat[Delay versus $N$ averaged over 10 rounds, $T_\mathrm{Tx} = 332 \mu$s ($K = 30$).]
	{\includegraphics[angle=0,width=0.43\textwidth]{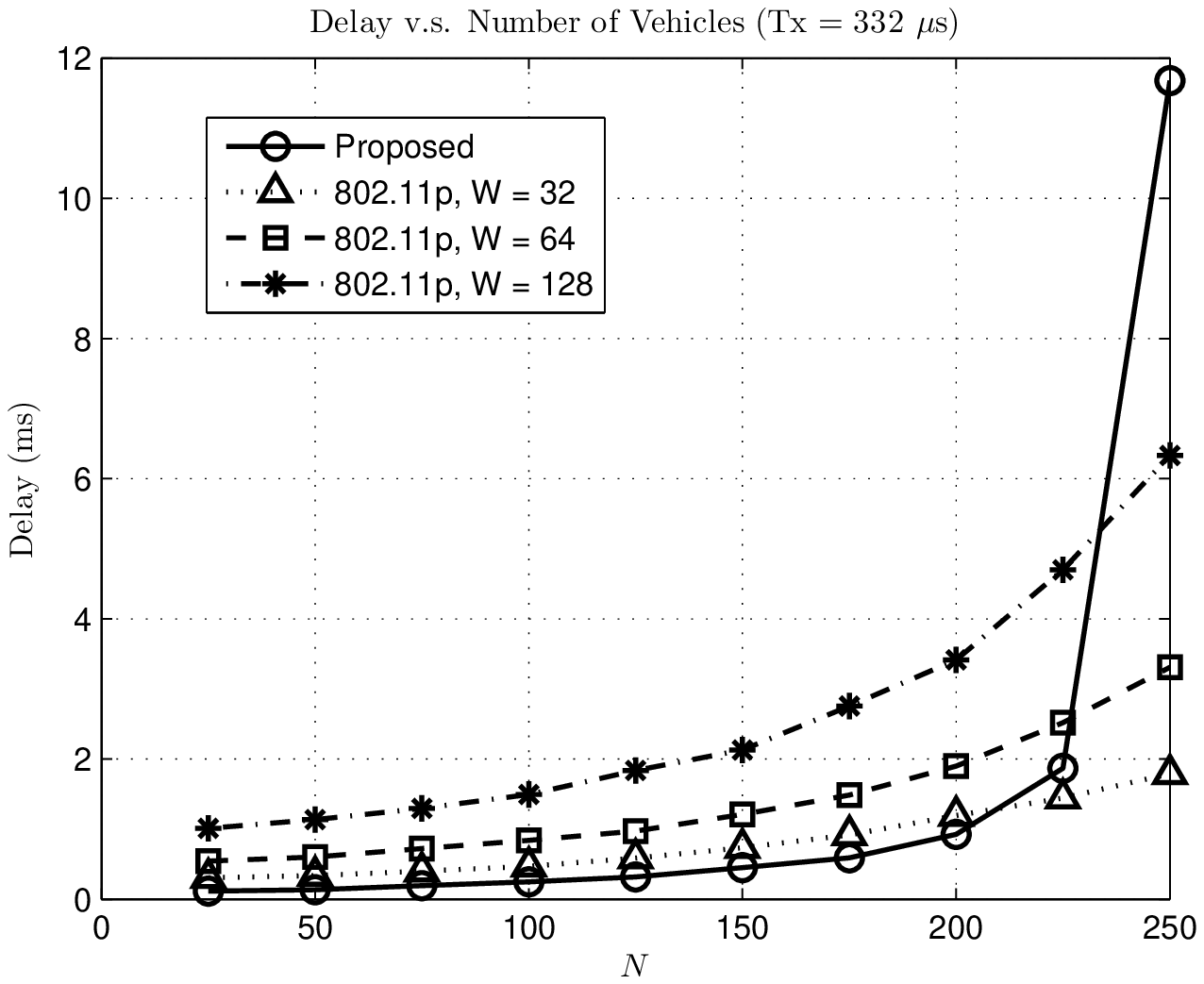}
		\label{f:Delay332}}
	\vspace{1mm}
	\caption{Delay versus number of vehicles under two different packet transmission time durations.}\label{f:Delay} \vspace{-5mm}
\end{figure}

Figs.~\ref{f:Delay254}~and~\ref{f:Delay332} demonstrate the average delay versus $N$ for $T_\mathrm{Tx} = 254 \mu$s and $T_\mathrm{Tx} = 332 \mu$s, respectively. In each plot, the delay of the CIDC is compared with that of 802.11p MAC with contention windows $W=32, W=64$, and $W=128$, respectively. From Fig.~\ref{f:Delay254}, it can be seen that the delay of the CIDC is smaller than that of 802.11p for all three contention window sizes $W$ and all $N$. From Fig.~\ref{f:Delay332}, it can be seen that the delay  performance of the CIDC is better at all points except for the cases of very large $N$, i.e., $N =225$ and $N=250$. For $N=225$, the delay of the CIDC is larger than that of 802.11p with $W = 32$ but smaller than that of 802.11p with $W=64$ or $W=128$. For $N=250$, the delay of the CIDC is the largest because the system is saturated and an increasing number of packets are contending for channel access under the CIDC. \textcolor{Black}{This reflects the fact that, compared with the 802.11p, the proposed design trades off delay for a smaller collision probability when $N$ is very large.} Nevertheless, it is worth noting that the case of a vehicle simultaneously exchanging messages with 250 other vehicles would be an extreme case in either an urban or a highway scenario. 

Comparing Figs.~\ref{f:Pcol}~and~\ref{f:Delay}, it can be observed that the CIDC has a better performance in terms of both collision probability and delay over a wide range of $N$ at the cost of a smaller saturation threshold in terms of $N$. Since the advantage of the CIDC holds for all $W$ in [32, 128], it follows that the contention intensity based MAC outperforms a design solely based on adapting the contention window $W$ of the 802.11p.

\subsection{A Comparison of Analytical and Numerical Results}

The second example compares the analytical results on the contention delay from Theorem~1 and the collision probability upper bound from Theorem~2 with the numerical results.  

\begin{figure}
	\begin{center}
		\includegraphics[scale = 0.54, angle = 0]{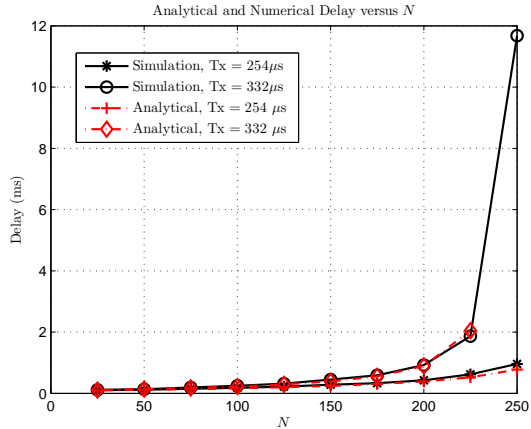}
		\caption{Comparison of the analytical and numerical results on the average contention delay versus $N$.}
		\label{f: AnaNumDelay}
	\end{center}
	\vspace{-4mm}
\end{figure}

Fig.~\ref{f: AnaNumDelay} demonstrates the analytical and numerical results on the average contention delay $d_\mathrm{c}$ versus $N$ for $T_\mathrm{Tx} = 254 \mu$s and $T_\mathrm{Tx} = 332 \mu$s. It can be seen that the analytical results match the numerical results. Moreover, the match is exact for small and medium $N$ while a small gap appears for large $N$. The reason is that the analytical results in Theorem~1 are for an unsaturated scenario while the system approaches saturation as $N$ increases. For the case $T_\mathrm{Tx} = 332 \mu$s, the system is beyond saturation at $N = 250$ and a solution cannot be found based on Theorem~1 anymore.

\begin{figure}[t]
	\centering \subfloat[Simulated collision probability and the analytical upper bound versus $N$, $T_\mathrm{Tx} = 254 \mu$s ($K = 24$).]
	{\includegraphics[angle=0,width=0.43\textwidth]{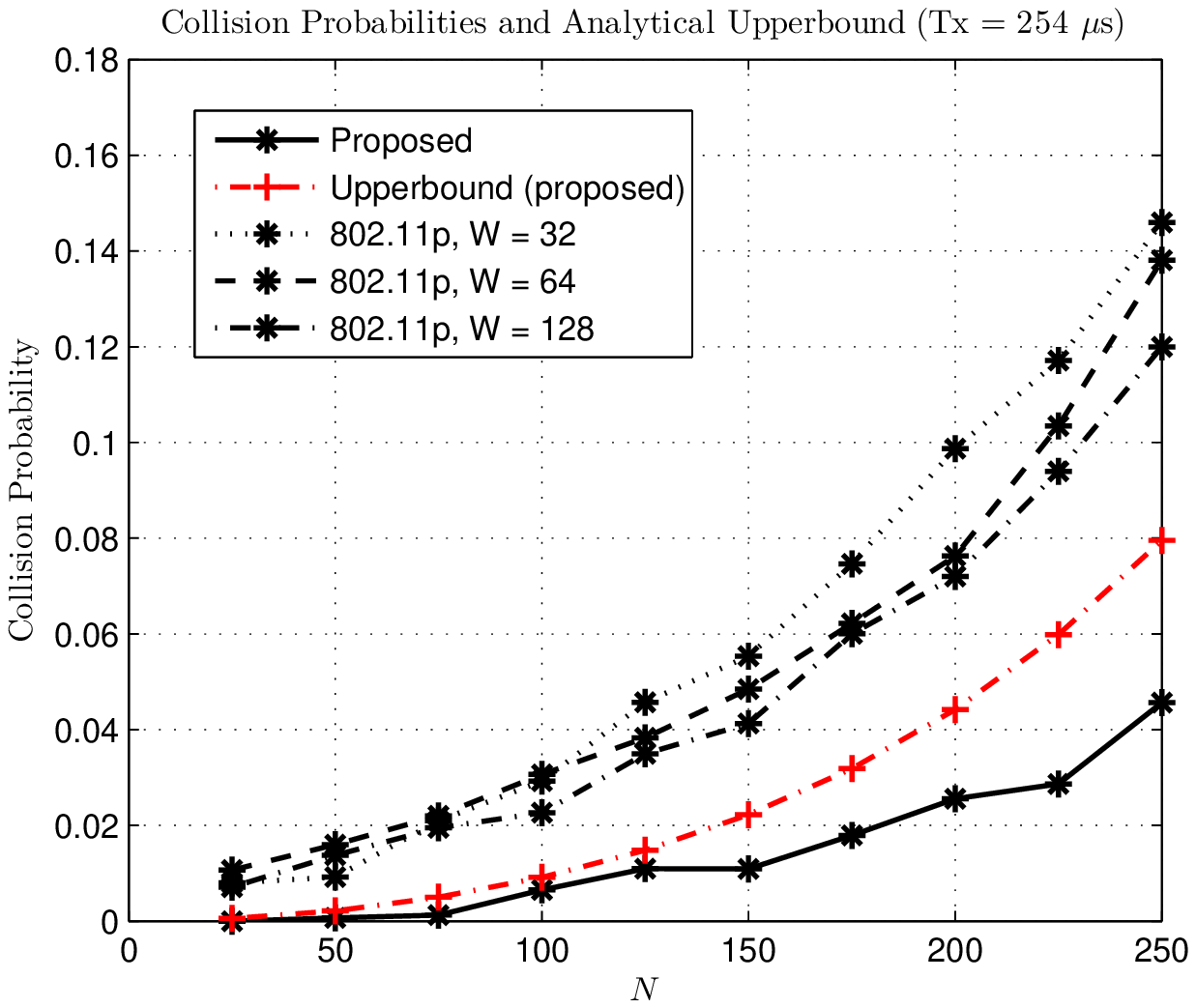}
		\label{f:AnaNumPcol254}}
	\hspace{1mm} 
	\subfloat[Simulated collision probability and the analytical upper bound versus $N$, $T_\mathrm{Tx} = 332 \mu$s ($K = 30$).]
	{\includegraphics[angle=0,width=0.43\textwidth]{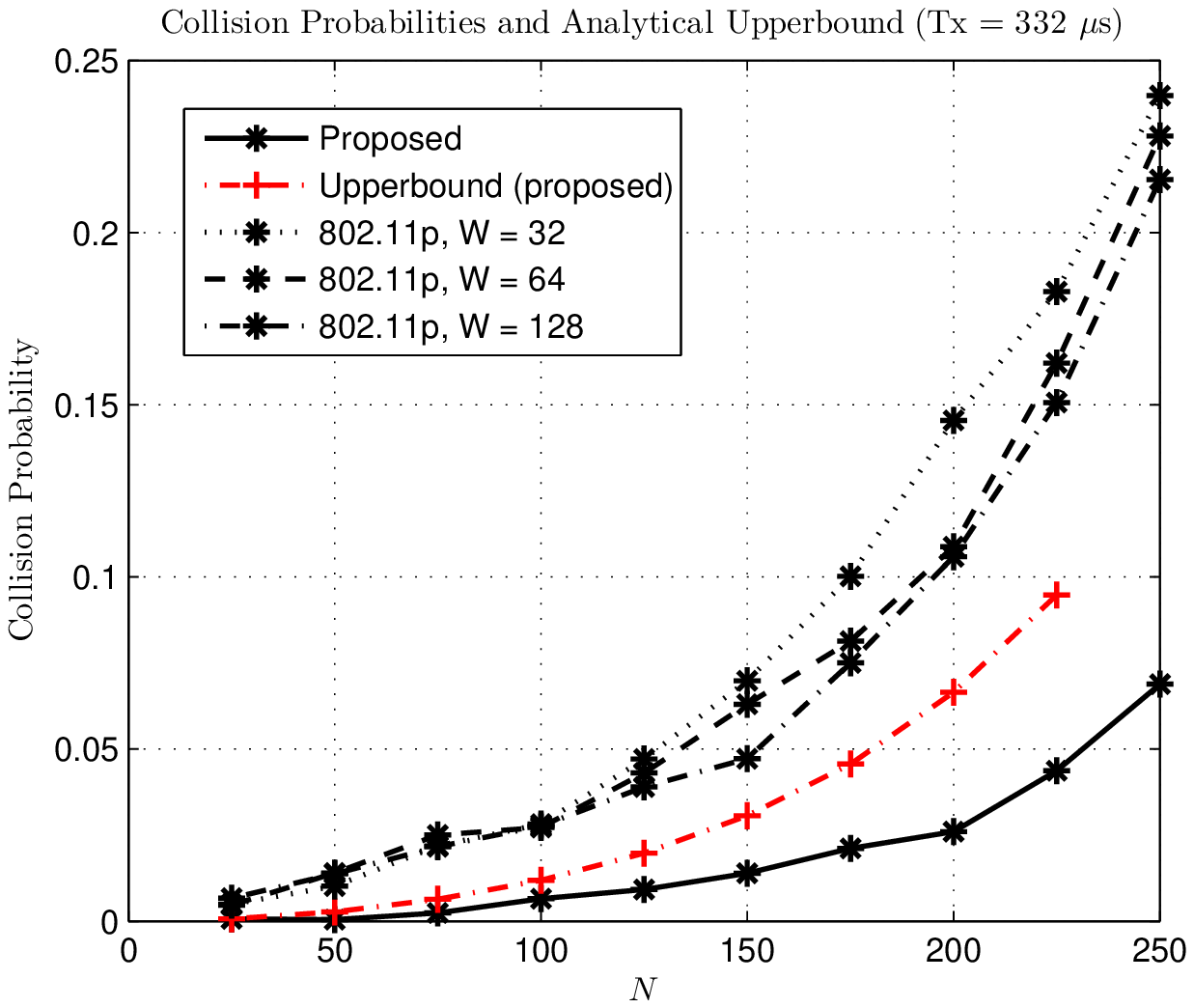}
		\label{f:AnaNumPcol332}}
	\vspace{1mm}
	\caption{Comparison of the analytical and numerical results on the collision probability versus $N$.}\label{f:AnaNumPcol} \vspace{-5mm}
\end{figure}

Figs.~\ref{f:AnaNumPcol254}~and~\ref{f:AnaNumPcol332} demonstrate the analytical upper bound of the collision probability and the numerical result on the collision probability versus $N$ for $T_\mathrm{Tx} = 254 \mu$s and $T_\mathrm{Tx} = 332 \mu$s, respectively. From the figures, it can be seen that the analytical upper bounds are valid although not very tight when $N$ is large. Nevertheless, even the upper bound in each figure lies below the collision probability of the 802.11p MAC for any $W$ in the entire range of $N$.

\subsection{Performance under Contention Intensity Estimation Errors}

The third example demonstrates the performance of the CIDC with estimation errors. The accurate estimation of the contention intensity based on information exchange and processing can be an ideal case. In practice, the estimation can be subject to errors due to the vehicle mobility. Consider a two-directional highway and vehicles moving at a speed of 80km/h on each side. Assume a communication range of 500m on both sides. Consider the extreme (worst) case that all other vehicles are moving towards the opposite direction of the target vehicle. The density of the vehicles on the other side of the highway is assumed to be constant. Then, the percentage of vehicles that leave and enter the communication range of the target vehicle is $0.89\%$ of $N$ within a message cycle of 100ms. Accordingly, in this simulation example, we use a parameter $\delta$ to represent that $\delta$ percent of the neighbor vehicles has changed in a message cycle. Instead of 0.89, which is already calculated from an extreme case, we further increase $\delta$ to $1$ and $3$ to accommodate a margin for other possibilities of errors. 

\begin{figure}[!t]
	\centering \subfloat[Collision probability versus $N$ with and without error in contention intensity estimation, $T_\mathrm{Tx} = 254 \mu$s ($K = 24$).]
	{\includegraphics[angle=0,width=0.43\textwidth]{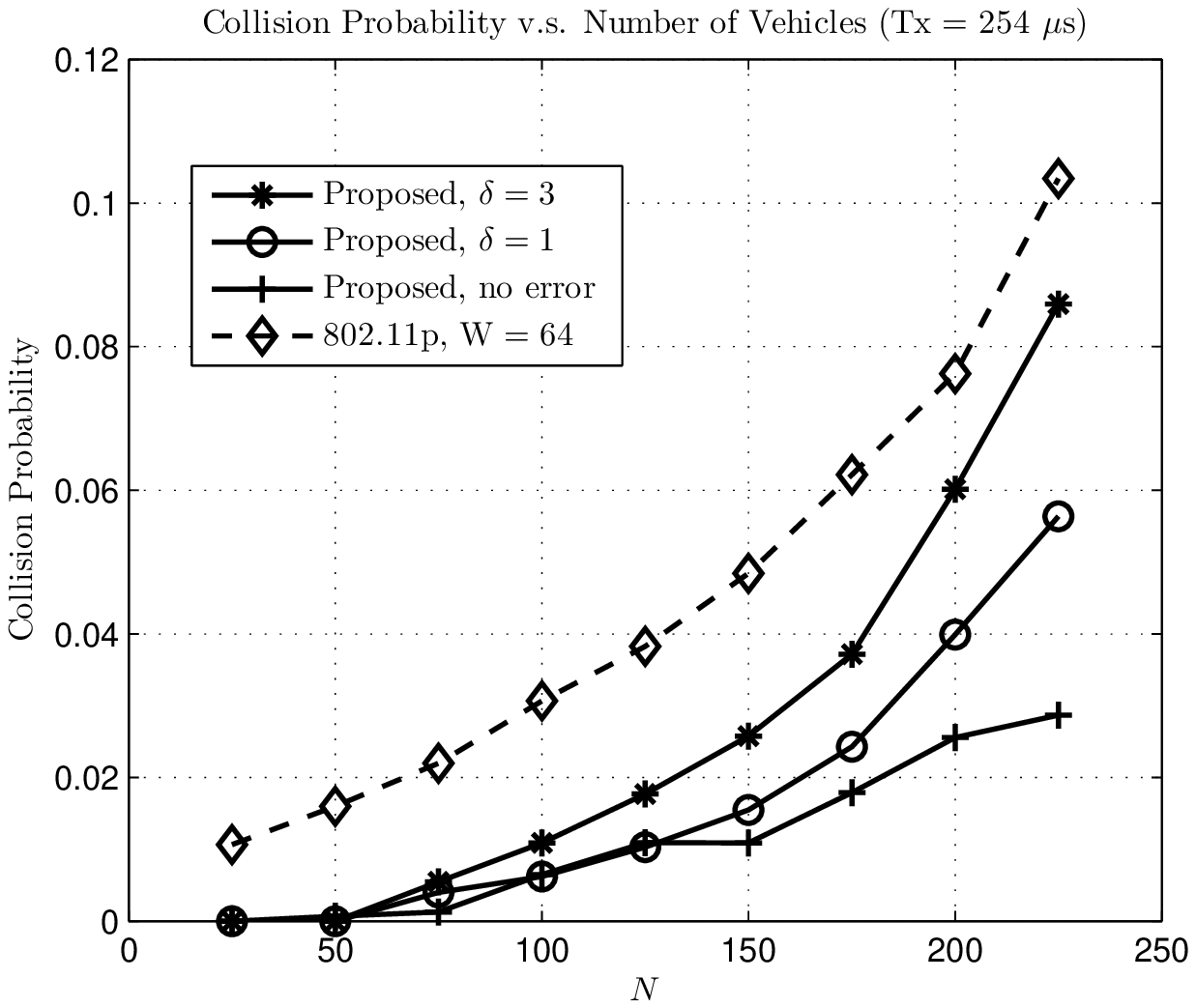}
		\label{f:Pcolerr}}
	\hspace{1mm} 
	\subfloat[Delay versus $N$ with and without error in contention intensity estimation, $T_\mathrm{Tx} = 254 \mu$s ($K = 24$).]
	{\includegraphics[angle=0,width=0.43\textwidth]{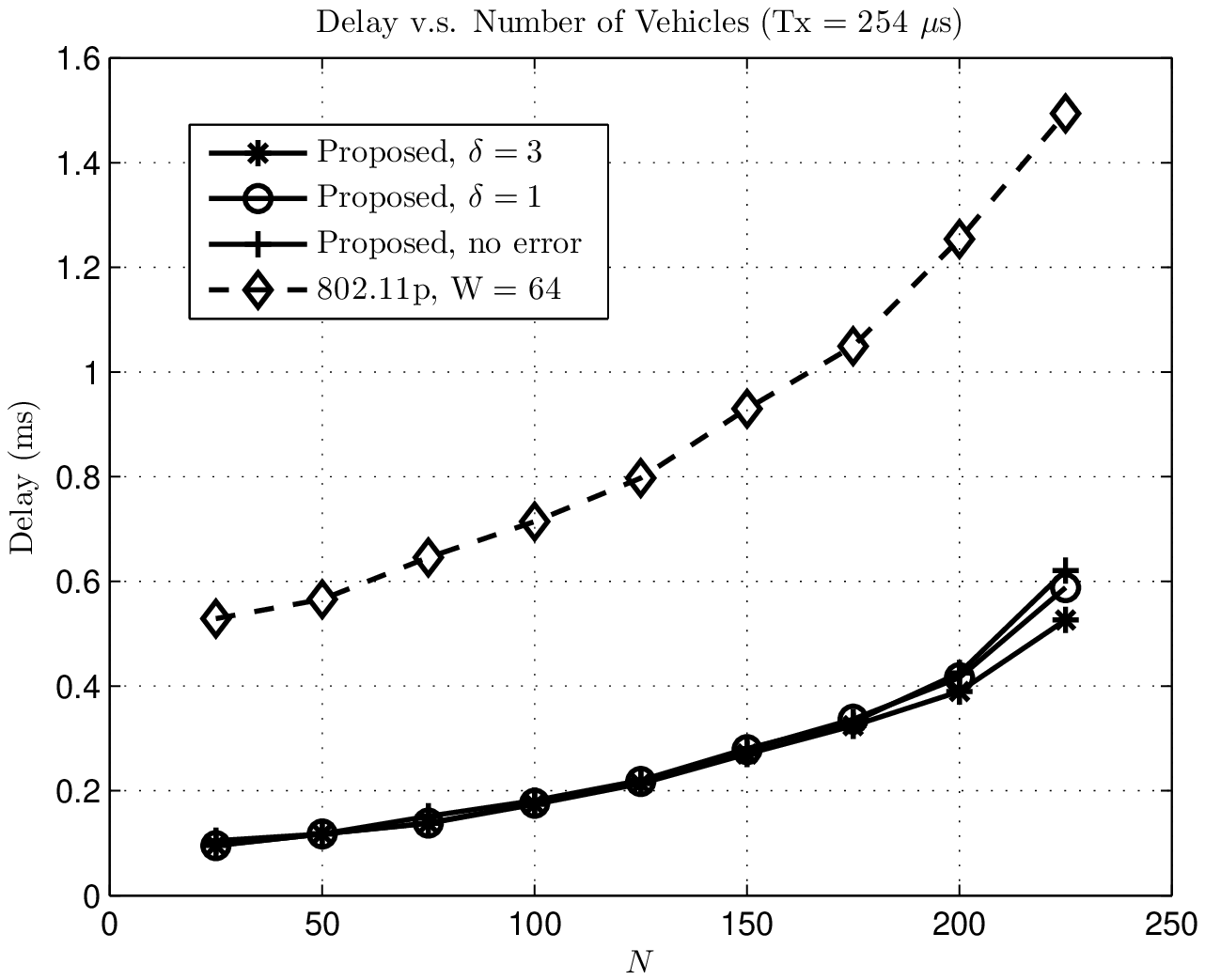}
		\label{f:Delayerr}}
	\vspace{1mm}
	\caption{Collision probability and contention delay with and without error in contention intensity estimation.}\label{f:PcolDelayerr} \vspace{-5mm}
\end{figure}

In Fig.~\ref{f:Pcolerr}, the collision probability of the CIDC with errors is compared to that of the CIDC without any error and the 802.11p MAC with $W=64$. It can be seen that introducing error in the estimation of contention intensity increases the collision probability. However, the performance is still better than 802.11p MAC with $W=64$ even for $\delta = 3$. The collision probability of the CIDC with error when $\delta = 1$ is much smaller than that of the 802.11p MAC. 

In Fig.~\ref{f:Delayerr}, the average contention delay of the CIDC with errors is compared to that of the CIDC without any error and the 802.11p MAC with $W=64$. It can be seen that the impact of estimation error on the contention delay is insignificant especially when $N$ is small. When $N$ is large, an estimation error slightly reduces the contention delay due to an increase in the collision probability.

\section{Conclusions}\label{s:con}

In this work, we have proposed a CIDC with application-layer and MAC-layer designs to improve the performance of safety message broadcast in V2V communications. Exploiting the unique features of the considered scenario, the development of the CIDC is an exploration of the performance-overhead trade-off and an effort to achieve a significant performance improvement with an overhead as small as possible in the scenario of safety message broadcast. As a result, the CIDC, which is distributed and compatible with 802.11p, substantially improves the performance of safety message broadcast at the cost of a small communication and computation overhead even when errors are introduced in the contention intensity estimation. With the above features, the CIDC is a promising candidate as either a building block in a comprehensive protocol for V2V communications or a general protocol that addresses periodical broadcast in a distributed network with a fixed or slow-varying topology.

\appendix

\subsection{Proof of Lemma~1}\label{s:ProofLemma1}

If $h(k) = 0$, it follows from \eqref{e:upsilonDef} that
\begin{align}
\upsilon(k[s]) &= \frac{c(k) + n_\mathrm{I}(k[1])}{\max\{b^{\max}(k), e_v(k[s])\}} \nonumber \\
&\leq \frac{c(k) + n_\mathrm{I}(k[1])}{e_v(k[s])} \leq \frac{c(k) + 1}{e_v(k[s])} = \frac{1}{M}.
\end{align}
Otherwise ($h(k) = 1$), it follows from \eqref{e:upsilonDef} that
\begin{align}
\upsilon(k[s]) &= \frac{c(k) + \sum\limits_{l = 1, l \in \mathcal{S}_k}^{s} n_\mathrm{I}(k[l])}{\max\{b^{\max}(k), e_v(k[s])\}} \nonumber \\
&\leq \frac{c(k) + \sum\limits_{l = 1, l \in \mathcal{S}_k}^{s} n_\mathrm{I}(k[l])}{e_v(k[s])} \nonumber \\
& \leq \frac{ c(k) + \sum\limits_{l=k_1, \in \mathcal{S}_k}^{s - 1} n_\mathrm{I}(k[l]) + 1 }{e_v(k[s])} = \frac{1}{M}.
\end{align}
This concludes the proof of Lemma~1.\hfill$\blacksquare$

\subsection{Proof of Lemma~2}\label{s:ProofLemma2}

The expected change of $c(k)$ in one slot is
\begin{align}
E\{\Delta c(k)\} = \left\{
\begin{array}{ll}
\lambda N T_\mathrm{s}, \quad\quad\quad\; \text{if} \;\; h(k) = 0 \\
\lambda N K T_\mathrm{s} - n_\mathrm{s}, \; \text{else}.
\end{array}
\right.
\end{align}
The probability of a slot being idle (i.e., $h(k) = 0$) and busy (i.e., $h(k) = 1$) are given by  
\begin{subequations}
\begin{align}
P(h(k) = 0) &= P_\mathrm{ck}(0) + (1 - P_\mathrm{ck}(0))(1 - \upsilon_\mathrm{s}) \\
P(h(k) = 1) &= (1 - P_\mathrm{ck}(0)) \upsilon_\mathrm{s}, 
\end{align}
\end{subequations}
respectively. Therefore, it must hold that
\begin{align}
(  P_\mathrm{ck}(0) + (1 - P_\mathrm{ck}(0))(1 - \upsilon_\mathrm{s}) ) \lambda N T_\mathrm{s} \nonumber \\ + (1 - P_\mathrm{ck}(0)) \upsilon_\mathrm{s} (\lambda N K T_\mathrm{s} - n_\mathrm{s}) =  0
\end{align}
in a steady state, which leads to \eqref{e:upsilonSteady}. \hfill$\blacksquare$

\subsection{Proof of Theorem~1}\label{s:ProofTheorem1}

Given the initial back-off counter selection rule \eqref{e:sysE2}, i.e, $e(k[s]) = (c(k[s])+ 1) M$, where
\begin{align}
c(k[s]) = c(k) + \!\! \sum\limits_{l =1, l \in \mathcal{S}_k}^{s-1} \! n_\mathrm{I}(k[l]),
\end{align}
the expected entry point in a steady state is 
\begin{align}
e_\mathrm{s} = \sum\limits_{j = 0}^{N} P_{c_k} (j) (j + 1 ) M = (c_\mathrm{s} + 1) M
\end{align}
where $P_\mathrm{c_k}(j) =  P(c(k[s]) = j)$. When there are $c(k[s]) = j > 1$ existing packets, the average number of transmissions a new packet endures (including its own) till the completion of its own transmission is $ 1 + j - 1/2$, in which $1$ refers to its own transmission and $-1/2$ corresponds to the fact that the first packet is half-way transmitted on average upon the arrival of the new packet.  Therefore, the delay of a packet due to busy slots (transmissions) is 
\begin{align}
d_\mathrm{o}^\mathrm{busy} & = \bigg(1 + \sum\limits_{j = 1}^{N} P_\mathrm{c_k} (j) (j - \frac{1}{2} ) \bigg) K T_\mathrm{s} \nonumber \\
& = \bigg(1 + c_\mathrm{s} -  \frac{1}{2}(1 - P_{c_k} (0))  \bigg) K T_\mathrm{s}.
\end{align}
Note that queue jumping, which can be viewed as a switch of transmission order, does not have an impact on the average number of transmissions a packet needs to wait for. Since the packet enters from $(c(k[s])+ 1) M$ and there are $c(k[s])$ busy slots, the average delay due to the number of empty slots is 
\begin{align}
d_\mathrm{o}^\mathrm{empty} = \bigg((c_\mathrm{s}  + 1)M - c_\mathrm{s}\bigg)T_\mathrm{s}.  
\end{align}  
The overall duration of contending for channel access and transmission of a packet is then
\begin{align}
d_\mathrm{o} = d_\mathrm{o}^\mathrm{busy} + d_\mathrm{o}^\mathrm{empty}, 
\end{align}
which leads to the equation \eqref{e:delayOverall}. 

In a steady state, $d_\mathrm{o} \leq 1/\lambda$ where strict inequality holds if the system is not saturated. As a packet arrives every $1/\lambda$ seconds, the probability that a vehicle has a contending packet is $P_\mathrm{con} = d_\mathrm{o}/(1/\lambda) = \lambda d_\mathrm{o}$. Given this individual contending probability, the expected number of contending packets should satisfy 
\begin{align}\label{e:Pcon}
N P_\mathrm{con} = c_\mathrm{s},
\end{align}
which leads to the equation \eqref{e:delayCs}. 

Given $P_\mathrm{con}$, the probability that no packet is contending for channel access is $(1 - P_\mathrm{con})^N$. Therefore, it must hold that  
\begin{align}\label{e:Pck0Pcon}
P_{c_k} (0) = (1 - P_\mathrm{con})^N. 
\end{align}
Substituting $P_\mathrm{con}$ from \eqref{e:Pcon} into \eqref{e:Pck0Pcon} gives the equation \eqref{e:delayPck0}. 

This completes the proof of Theorem~1. \hfill $\blacksquare$

\subsection{Proof of Lemma~3}\label{s:ProofLemma3}

From \eqref{e:delayOverall} to \eqref{e:delayPck0} in Theorem~1, $c_\mathrm{s}$ must satisfy
\begin{align}
&c_\mathrm{s} \bigg( 1 - N \lambda T_\mathrm{s} (K + M -1) \bigg) \nonumber \\
& \qquad= N \lambda \bigg( K + M - \frac{K}{2} \bigg(1 - \big(1 -\frac{c_\mathrm{s}}{N}\big)^N\bigg)\bigg)  T_\mathrm{s}.
\end{align}
For a small $N$, $(1 - c_\mathrm{s}/N)^N$ can be approximated by 1 and it gives the result \eqref{e:smallNcs}. For a large $N$, $(1 - c_\mathrm{s}/N)^N$ approaches 0 and it gives the result \eqref{e:largeNcs}.  \hfill $\blacksquare$

\subsection{Proof of Lemma~4}\label{s:ProofLemma4}
Using \eqref{e:sysE2}, the entry point for packets $A$ and $B$ are
\begin{subequations}
	\begin{align}
	e_A(k_1[s_1]) = M\bigg(c(k_1) + \sum\limits_{l =1}^{s_1} \! n_\mathrm{I}(k_1[l])\bigg) \\
	e_B(k_2[s_2]) = M\bigg(c(k_2) + \sum\limits_{l =1}^{s_2} \! n_\mathrm{I}(k_2[l])\bigg).
	\end{align}
\end{subequations}
The back-off counter of $A$ decreases by $\alpha$ from slot $k_1$ to slot $k_2$. Therefore, for packets $A$ and $B$ to collide, it must hold
\begin{align}
M\bigg(c(k_1) \!+\! \sum\limits_{l =1}^{s_1} \! n_\mathrm{I}(k_1[l])\bigg) \! - \!\alpha = M\bigg(c(k_2) \!+ \! \sum\limits_{l =1}^{s_2} \! n_\mathrm{I}(k_2[l])\bigg)
\end{align}
Based on the number of transmitted and arrived packets, it holds that
\begin{align}
c(k_2) + \sum\limits_{l =1}^{s_2}n_\mathrm{I}(k_2[l]) = c(k_1) + \sum\limits_{l =1}^{s_1} \! n_\mathrm{I}(k_1[l])  + \eta - \tau.
\end{align}
Therefore,  packets $A$ and $B$ collide only if $M (\tau - \eta)= \alpha$.  \hfill $\blacksquare$

\subsection{Proof of Theorem~2}\label{s:ProofTheorem2}

Unlike the packet-perspective collision analysis for 802.11p, the proof for the proposed model is based on a slot perspective. Consider an arbitrary slot $k$,  a collision is determined (and will happen in a later slot) in this slot under the following necessary conditions:
\begin{itemize}
	\item [i.] there is at least one contending packet, i.e., $c(k) > 0$.   
	\item [ii.] the virtual entry point at the current mini-slot $e_\upsilon(k[s])$ corresponds to a busy slot, i.e., $h(e_\upsilon(k[s])) = 1$. 
	\item [iii.] there is a new packet arrival in slot $k$, i.e., $n_\mathrm{I}(k)>0$. 	
\end{itemize} 
The probabilities that the above conditions are satisfied are given as follows  
\begin{itemize}
	\item [i.] $1 - P_\mathrm{c_k}(0)$
	\item [ii.] 0 when $e_\upsilon(k[s]) > b^{\max}(k)$ and $\upsilon_\mathrm{s}$ when $e_\upsilon(k[s]) < b^{\max}(k)$.   
	\item [iii.] $1 - (1 - \lambda T_\mathrm{s})^N $ if slot $k$ is idle and $ 1 - (1 - \lambda K T_\mathrm{s})^N $ if slot $k$ is busy. 	
\end{itemize} 
Therefore, the probability that a collision is determined in slot $k$ (and will happen in a later slot) is upper-bounded by
\begin{align}
P_\text{I}^\text{s} = \upsilon_\mathrm{s} (1 - P_\mathrm{c_k}(0))   \bigg(1 - (1 - \lambda T_\mathrm{s})^N \bigg)
\end{align}
in an idle slot and 
\begin{align}
P_\text{B}^\text{s} =   \upsilon_\mathrm{s}  (1 - P_\mathrm{c_k}(0))   \bigg(1 - (1 - \lambda K T_\mathrm{s})^N \bigg)
\end{align}
in a busy slot. Since the packet-to-slot ratio is $\upsilon_\mathrm{s}$, each packet corresponds to  $1/\upsilon_\mathrm{s}$ equivalent exclusive slots on average, with 1 busy slots and $1/\upsilon_\mathrm{s} -1$ idle slots. Therefore, the collision probability for a packet is upper-bounded by
\begin{align}
P_\text{col}^\text{UB} = P_\text{B}^\text{s}  + \bigg(\frac{1}{\upsilon_\mathrm{s}} -1\bigg) P_\text{I}^\text{s} = \upsilon_\mathrm{s} (a_K - a_1) + a_1.
\end{align}
Using the result \eqref{e:ns2}, it holds that
\begin{align}
n_\mathrm{s} - 1 < \upsilon_\mathrm{s} (a_K - a_1) + a_1,
\end{align}
Substituting \eqref{e:upsilonSteady} into the above equation, it holds that
\begin{align}
n_\mathrm{s} - 1 < \frac{a_K - a_1}{1- P_\mathrm{ck}(0)} \frac{\lambda N T_\mathrm{s}}{n_\mathrm{s} - \lambda N (K-1)T_\mathrm{s}} + a_1.
\end{align}
Solving the above inequality and using \eqref{e:ns2} again gives \eqref{e:PcolUB}.\hfill $\blacksquare$

%
\medskip

\balance

\end{document}